\documentclass[12pt,twocolumn]{aastex62}
\usepackage{graphics}
\usepackage{color}
\usepackage{amsmath}
\usepackage{empheq} 
\newcommand{\ltsimeq}{\la}
\newcommand{\gtsimeq}{\ga}

\newcommand{\msun}{M$_{\odot}$}
\newcommand{\zsun}{Z$_{\odot}$}
\newcommand{\hi}{H{\sc i}}
\newcommand{\hii}{H{\sc ii}}
\newcommand{\ha}{H$\alpha$}

\shortauthors{McQuinn et al.}
\shorttitle{The Leoncino Dwarf Galaxy: the Low-Metallicity End of the LZ and MZ Relations}

\begin{document}
\title{THE LEONCINO DWARF GALAXY: EXPLORING THE LOW-METALLICITY END OF THE LUMINOSITY-METALLICITY AND MASS-METALLICITY RELATIONS\footnote{Based on observations made with the NASA/ESA Hubble Space Telescope, obtained at the Space Telescope Science Institute, which is operated by the Association of Universities for Research in Astronomy, Inc., under NASA contract NAS 5-26555. These observations are associated with program HST-GO-15243.}}

\author{Kristen.~B.~W. McQuinn}
\affiliation{Rutgers University, Department of Physics and Astronomy, 136 Frelinghuysen Road, Piscataway, NJ 08854, USA} 
\email{kristen.mcquinn@rutgers.edu}

\author{Danielle A.~Berg}
\affiliation{Department of Astronomy, The Ohio State University, 140 West 18th Avenue, Columbus, OH 43210, USA}

\author{Evan D. Skillman},
\affiliation{University of Minnesota, Minnesota Institute for Astrophysics, School of Physics and Astronomy, 116 Church Street, S.E., Minneapolis, MN 55455, USA} 

\author{Elizabeth Adams}
\affiliation{ASTRON, The Netherlands Institute for Radio Astronomy, Oude Hoogeveensedijk 4, 7991 PD, Dwingeloo, The Netherlands}
\affiliation{Kapteyn Astronomical Institute, University of Groningen Postbus 800, 9700 AV Groningen, The Netherlands}

\author{John M. Cannon}
\affiliation{Department of Physics and Astronomy, Macalester College, Saint Paul, MN 55105, USA}

\author{Andrew E.~Dolphin}
\affiliation{Raytheon Company, 1151 E. Hermans Road, Tucson, AZ 85756, USA}
\affiliation{University of Arizona, Steward Observatory, 933 North Cherry Avenue, Tucson, AZ 85721, USA}

\author{John J. Salzer}
\affiliation{Department of Astronomy, Indiana University, 727 East Third Street, Bloomington, IN 47405, USA}

\author{Riccardo Giovanelli}
\affiliation{Center for Astrophysics and Planetary Science, Space Sciences Building, Cornell University, Ithaca,
NY 14853, USA}

\author{Martha P.~Haynes}
\affiliation{Center for Astrophysics and Planetary Science, Space Sciences Building, Cornell University, Ithaca,
NY 14853, USA}

\author{Alec S. Hirschauer}
\affiliation{Space Telescope Science Institute, Baltimore, MD USA}

\author{Steven Janoweicki}
\affiliation{University of Texas at Austin, McDonald Observatory, 2515 Speedway, Stop C1400 Austin, Texas 78712, USA}

\author{Myles Klapkowski}
\affiliation{Department of Physics and Astronomy, Macalester College, Saint Paul, MN 55105, USA}

\author{Katherine L.~Rhode}
\affiliation{Department of Astronomy, Indiana University, 727 East Third Street, Bloomington, IN 47405, USA}

\begin{abstract}
Extremely metal-poor (XMP) galaxies are low-mass, star-forming galaxies with gas-phase oxygen abundances below 12$+$log(O/H) $=$ 7.35 ($\sim\frac{1}{20}$ \zsun). Galaxy evolution scenarios suggest three pathways to form an XMP: (1) secular evolution at low galaxy masses, (2) slow evolution in voids, or (3) dilution of measured abundances from infall of pristine gas. The recently discovered XMP galaxy Leoncino, with an oxygen abundance below 3\% \zsun, provides an opportunity to explore these different scenarios. Using Hubble Space Telescope imaging of the resolved stellar populations of Leoncino, we measure the distance to the galaxy to be D$=12.1^{+1.7}_{-3.4}$ Mpc and find that Leoncino is located in an under-dense environment. Leoncino has a compact morphology, hosts a population of young, massive stars, has a high gas-to-star mass ratio, and shows signs of interaction with a galaxy nearby on the sky, UGC~5186. Similar to nearly all XMP galaxies known in the nearby universe, Leoncino is offset from the Luminosity-Metallicity (LZ) relation. Yet, Leoncino {\it is consistent} with the stellar Mass-Metallicity (MZ) relation defined by Local Volume galaxies. Thus, our results suggest that the offset from the LZ relation is due to higher recent star formation, likely triggered by a minor interaction, while the low oxygen abundance is consistent with the expectation that low-mass galaxies will undergo secular evolution marked by inefficient star formation and metal-loss via galactic winds. This is in contrast to XMP galaxies that are outliers in {\it both} the LZ and MZ relations; in such cases, the low oxygen abundances are best explained by dilution due to the infall of pristine gas. We also discuss why quiescent XMP galaxies are underrepresented in current surveys.

\end{abstract} 

\section{Chemical Evolution Pathways for Metal-Poor, Star-Forming Galaxies}\label{sec:intro}
The gas-phase oxygen abundance of galaxies is known to correlate with stellar mass or, as a proxy for stellar mass, luminosity. A strong correlation between luminosity and metallicity was seen in early studies \citep[e.g., the LZ relation;][]{Skillman1989a}. Later studies focused on the tighter relation between stellar mass and metallicity (the MZ relation) where oxygen abundances increase with mass for low-mass systems \citep[$M_* \ltsimeq 10^9$ \msun;][]{Berg2012} and begin to flatten for massive spiral galaxies ($M_* \gtsimeq 10^{10.5}$ \msun; \citealt{Tremonti2004}; but see also \citealt{Hirschauer2018}). The MZ relation defined by nearby galaxies is thought to arise from the combined impact of star formation, star formation efficiency, metal loss through outflows, and, to a lesser extent, the dilution of abundances by the infall of pristine gas \citep[e.g.,][]{Dalcanton2007}. Thus, while the interplay of these various factors is complex and can vary for any given system, the location of a galaxy in the MZ plane provides constraints for the cumulative chemical evolution of that object.

Galaxies at the low-metallicity end of the MZ relation are of particular interest. Systems with abundances below 12$+$log(O/H) $=$ 7.65, corresponding to roughly $\frac{1}{10}$ the solar oxygen abundance,\footnote{\citet{Kunth2000} adopted a solar oxygen abundance of 12 $+$ log(O/H) $= 8.91$. Here we adopt a more modern value from \citet{Asplund2009} of 8.69.} have been previously classified as ``very metal deficient'' galaxies \citep[e.g.,][]{Kunth2000}. This early definition was based on the O/H distribution of star-forming dwarf galaxies which showed a peak at $\sim\frac{1}{10}$ \zsun\ with a sharp drop below that value. Recently, there have been a growing number of galaxies discovered below this threshold \citep[e.g.,][]{Izotov2006, Ekta2008, Brown2008, Papaderos2008, Izotov2012, Skillman2013, Berg2016, Yang2017, Hsyu2017, James2017, Guseva2017, Hirschauer2016, Izotov2019, Hsyu2018, Berg2019, Senchyna2019a, Senchyna2019b} and as low as 12$+$ log(O/H) $=$ 6.98 \citep{Izotov2018}. Therefore, to isolate the most extreme systems, we classify galaxies as ``extremely metal-poor'', or XMP, with the formal definition of  having an oxygen abundance equal to or below 12$+$ log(O/H) $\leq$ 7.35, which is roughly $\frac{1}{20}$ the solar oxygen abundance. We are choosing a fixed O/H ratio, instead of a fixed percentage of the solar oxygen abundance, as the derived values of the solar abundance continues to change with time \citep[e.g.,][]{Anders1989, Grevesse1998, AllendePrieto2001, Asplund2004, Asplund2009, Scott2009}. 

XMP galaxies provide boundary conditions and essential constraints on chemical-enrichment pathways. Thus, in an era when galaxy formation simulations are able to reproduce the MZ relation at increasingly low masses \citep[e.g.,][]{Ma2016}, XMP galaxies have a unique role to play in constraining simulations. XMP galaxies are also laboratories for studying the formation and evolution of massive stars in nearly pristine gas \citep[e.g.,][]{Garcia2019}, and for providing constraints on the primordial helium abundance \citep[[e.g.,][]{Izotov1994, Stasinska2001, Skillman2013, Cooke2015, Aver2015}. In addition, XMP galaxies that can be studied locally offer an opportunity to understand the details of star formation and chemical evolution in a regime analogous to that of chemically primitive galaxies in the early universe. This is particularly relevant given that upcoming JWST observations at high redshift have the potential for discovering primeval galaxies at similar metallicities, but with limited ability to study the systems in detail.

Early searches for metal-poor galaxies discovered a few of these extreme systems with oxygen abundances below 5\% \zsun, including the iconic blue compact dwarf galaxies I~Zw~18, SBS~0335-052W, and DDO~68, with robust oxygen abundances of 12$+$log(O/H) $=\ 7.17\pm0.04, 7.12\pm0.03, 7.21\pm0.03$, respectively \citep{Skillman1993, Izotov2005, Pustilnik2005}. These systems are all high surface brightness galaxies that are significantly offset from expectations of the LZ relation for typical, late-type galaxies.

Using the properties of these galaxies as a guide, searches for more XMP galaxies have been carried out with varying success. Though progress was initially slow, as listed above, the success rate in finding XMP galaxies has increased within the last few years. Nearly all of the newly discovered XMP galaxies are outliers on the LZ relation. This is best understood by extrapolating what is known from the iconic XMP galaxies, namely that the paucity of metals may be due to dilution of the gas-phase metallicity by pristine gas falling into the galaxy from the outer disk or local environment, likely triggered by an interaction. In this scenario, the infalling gas quickly mixes into the interstellar media (ISM) and lowers the measured abundance, while simultaneously producing a significant burst of star formation and correspondingly, an increase in the galaxy luminosity \citep{Ekta2010b}. The relative infrequency of such events provides a natural explanation for the paucity of such systems.

Alternatively, environment has been suggested to play a role. In a study of dwarf galaxies in low-density void environments, low-mass galaxies were found to lie below the LZ relation in a parallel sequence. Their gas-phase oxygen abundances were measured to be 30$-$40\% lower on average than similar galaxies in typical field environments \citep{Pustilnik2016, Kniazev2018}. The authors have suggested that star formation is less efficient in voids, resulting in a slower galaxy evolution process. However, the oxygen abundances were measured by a combination of the direct method (utilizing the temperature sensitive [O III] $\lambda4363$ auroral line), and semi-empirical and empirical methods using strong emission line ratios. These have been found to differ by as much as a few tenths of dex and could introduce a bias when comparing abundance measurements \citep[e.g.,][]{vanZee2006, Kewley2008, Moustakas2010, Andrews2013}. If direct-method oxygen abundances alone are considered, the differences between galaxies in voids and more typical galaxies in the Local Volume are minimized (see analysis in \S\ref{sec:lz} below). 

Interestingly, there is an XMP galaxy that {\em does} agree with both the MZ and the LZ relation. A novel breakthrough for XMP galaxy searches came with the discovery of Leo~P, a low-luminosity, star-forming, extremely low-mass galaxy with an oxygen abundance comparable to I~Zw~18 \citep[7.17$\pm$0.04;][]{Skillman2013}. Leo~P was detected via its gas content in the Arecibo Legacy Fast ALFA blind \hi\ survey  \citep[ALFALFA;][]{Giovanelli2005, Haynes2011, Giovanelli2013, Haynes2018} and confirmed as a galaxy with follow-up optical imaging \citep{Rhode2013}. The paucity of metals in Leo~P is consistent with expectations from secular evolution processes (i.e., from inefficient star formation and galactic winds). Based on the star-formation and chemical-evolution history of the galaxy derived from the resolved stellar populations, Leo~P has lost 95$\pm2$\% of its oxygen, likely via stellar-feedback driven galactic winds \citep{McQuinn2015e, McQuinn2015f}. 

Through the ALFALFA survey, we have identified another XMP galaxy with an oxygen abundance even lower than Leo~P (see \S\ref{sec:properties}). Colloquially referred to as the Leoncino Dwarf (AGC~198691), spectroscopic observations of the \hii\ region yield a gas-phase oxygen abundance of 12$+$log(O/H) $=$ 7.12$\pm0.04$,\footnote{Updated value based on new, deep, Large Binocular Telescope spectra (E.~Aver et al., in preparation), 0.1 dex higher than the originally reported oxygen abundance of 7.02$\pm0.03$ in \citet{Hirschauer2016}.} or less than 3\% \zsun. This is equivalent to the oxygen abundance in SBS 0335-052W and less than that measured in both I~Zw~18 and DDO~68. 

\begin{table}
\begin{center}
\caption{Properties of the Leoncino Dwarf Galaxy}
\label{tab:properties}
\end{center}
\begin{center}
\vspace{-10pt}
\begin{tabular}{lr}
\hline 
\hline
Parameter			& Value	\\
\hline
R.A. (J2000)		& 9:43:32.4	\\
Decl. (J2000)		& $+$33:26:58.0	\\
12$+$log(O/H) 		& 7.12$\pm0.04$ \\
V$_{\rm helio}$		& 514 km s$^{-1}$ \\
V$_{\rm gsr}$		& 481 km s$^{-1}$ \\
M$_{\rm HI}$ / M$_*$& 25 \\
m$_V$			& 19.53 $\pm0.03$ mag \\
F$_{3.6~\micron}$ 	& (1.50$\pm$0.07) $\times10^{-5}$ Jy \\
m$_{3.6~\micron}$	& 18.8 mag \\
$B-V$ 			& 0.29$\pm0.04$ mag \\
$A_V$	 		& 0.04 mag\\
P.A.				& 80$^{\circ}$ \\
semi-major axis		& 6.\arcsec75\\
eccentricity		& 0.61 \\
WFC3 F606W exp. time & 15018 s \\
WFC3 F814W exp. time & 18618 s \\
\hline\\
\multicolumn{2}{c}{Distance-Dependent Values} \\
\hline
\hline
Parameter		& Value	\\
\hline
\vspace{2pt}
Distance Modulus& $30.4^{+0.31}_{-0.60}$ mag\\
\vspace{2pt}
Distance		&12.1$^{+1.7}_{-3.4}$ Mpc\\
\vspace{2pt}
SGX			& $4.6^{+1.3}_{-0.6}$ Mpc \\
\vspace{2pt}
SGY			& $9.8^{+2.7}_{-1.4}$  Mpc\\
\vspace{2pt}
SGZ			& $-5.4^{+1.5}_{-0.8}$ Mpc \\
\vspace{2pt}
Major-axis		& 0.8 kpc \\
$\vspace{2pt}
M_B$		& $-10.63^{+0.31}_{-0.60}$ mag \\
\vspace{2pt}
M$_{3.6~\micron}$& $-12.23^{+0.31}_{-0.60}$ mag \\
\vspace{2pt}
M$_*$  		& $(7.3^{+2.2}_{-4.3})\times10^5$ \msun\\ 
M$_{\rm HI}$ 	& $1.83\times10^7$ \msun \\
\hline
\end{tabular}
\end{center}
\tablecomments{Summary of the properties of Leoncino reported in this work and in \citet{Hirschauer2016} with an updated oxygen abundance from E.~Aver et al.\ (in preparation). V$_{\rm helio}$ and V$_{\rm gsr}$ are the heliocentric velocity and velocity with respect to the Galactic standard of rest respectively. $A_V$ is the Galactic extinction from \citet{Schlafly2011}. SGX, SGY, SGZ are the Supergalactic coordinates of Leoncino including uncertainties. The major-axis is the diameter of the galaxy in kpc based on the angular radius of 6.\arcsec75 and adopting our distance. $M_B$ is the extinction corrected, absolute $B$-band luminosity of the galaxy. M$_*$ is the stellar mass determined from IRAC [3.6] fluxes and assuming a M/L of 0.47 (J.~M.~Cannon et al., in preparation). See text for details.}
\vspace{1pt}
\end{table}

Here, using newly-obtained Hubble Space Telescope ($HST$) imaging to constrain the distance to Leoncino, we investigate the properties and environment of the system, use the distance to place the galaxy securely on the LZ and MZ relations, and explore different pathways of becoming an XMP in the nearby universe.  

We present a summary of the known properties of Leoncino in \S\ref{sec:properties}, the observations and data reduction in \S\ref{sec:obs}, and our distance measurement in \S\ref{sec:distance}. We explore the location of Leoncino in the nearby universe, both spatially and as a function of radial velocity in \S\ref{sec:env}. We examine the location of Leoncino in the LZ relation with other typical star-forming dwarfs in the field and star-forming dwarfs in voids. We also compare Leoncino with other XMP galaxies in both the LZ and MZ relations in \S\ref{sec:lz}. We discuss the different chemical-evolution pathways to form such extremely low-metallicity systems, their common characteristics, and the rarity of finding low-luminosity XMP galaxies in \S\ref{sec:discuss}. We summarize our conclusions in \S\ref{sec:conclusions}. 

\section{Properties of Leoncino from a Coordinated Observing Campaign}\label{sec:properties}
Leoncino was discovered via its \hi\ line emission in ALFALFA. Initial gas-mass estimates using a velocity-based distance from flow models suggested that it was very low mass ($M_{HI} \ltsimeq 10^7$ \msun) and, thus, was included in the Survey of \hi\ in Extremely Low-mass Dwarfs \citep[SHIELD;][]{Cannon2011c}, a follow-up study of Local Volume ALFALFA galaxies with \hi\ masses below 10$^{7.2}$ \msun. 

As part of the SHIELD program, we carried out a coordinated effort to characterize the chemical, stellar, and gas content of Leoncino; a summary of these properties is presented in Table~\ref{tab:properties}. Ground-based optical imaging from the WIYN 0.9m telescope in $R$-band and \ha\ confirmed the presence of the stellar component in the galaxy and an \hii\ region. Optical spectroscopy of the \hii\ region from the KPNO 4.0m and the MMT 6.5m telescopes showed that the galaxy was extremely metal-poor \citep{Hirschauer2016}. Recently obtained Large Binocular Telescope optical spectroscopy enabled a new measurement of the temperature-sensitive [O III] $\lambda$4363 auroral line, yielding a ``direct method'' gas-phase oxygen abundance of 12$+$log(O/H)$= 7.12 \pm 0.04$ (E.~Aver et al., in preparation). 

Combining the ALFALFA \hi\ data with a stellar mass determined from an infrared mass-to-light ratio (see \S\ref{sec:env}), yields a distance-independent \hi-to-stellar mass ratio of 25.\footnote{The original \hi-to-stellar mass ratio estimate of 50 from \citet{Hirschauer2016} used a stellar mass estimate from spectral energy distribution fitting which was deemed highly uncertain. Using the newly obtained IRAC imaging, this value is revised downward to 25.} This value is significantly higher than the typical value of $\sim1$ for dwarf irregulars, but less than a third of the mass ratio of 81 found in the extremely low-surface brightness galaxy Coma~P, also discovered by ALFALFA \citep{Brunker2019}. \hi\ observations from the Jansky Very Large Array (JVLA) map the \hi\ distribution with higher spatial resolution, showing the \hi\ extends much farther than the stellar distribution (see \S\ref{sec:env} and Figure~\ref{fig:hi}). The \hi\ kinematics of Leoncino from the JVLA data are complicated, with no coherent velocity gradient (J.~M.~Cannon et al., in preparation). 

From the survey of SHIELD galaxies, single-orbit $HST$ Advanced Camera for Surveys \citep[ACS;][]{Ford1998} imaging of the resolved stars in Leoncino (GO-13750, PI Cannon; K.~B.~W.~McQuinn et al., in preparation) revealed that the galaxy is highly compact with a crowded inner field and contains both a young, blue stellar population and an underlying older, red stellar population. This initial imaging confirmed that the galaxy is near enough for $HST$ to resolve the stars into individual point sources, but the data are of insufficient depth to provide a constraint on the distance. 

\begin{figure*}
\includegraphics[width=0.49\textwidth]{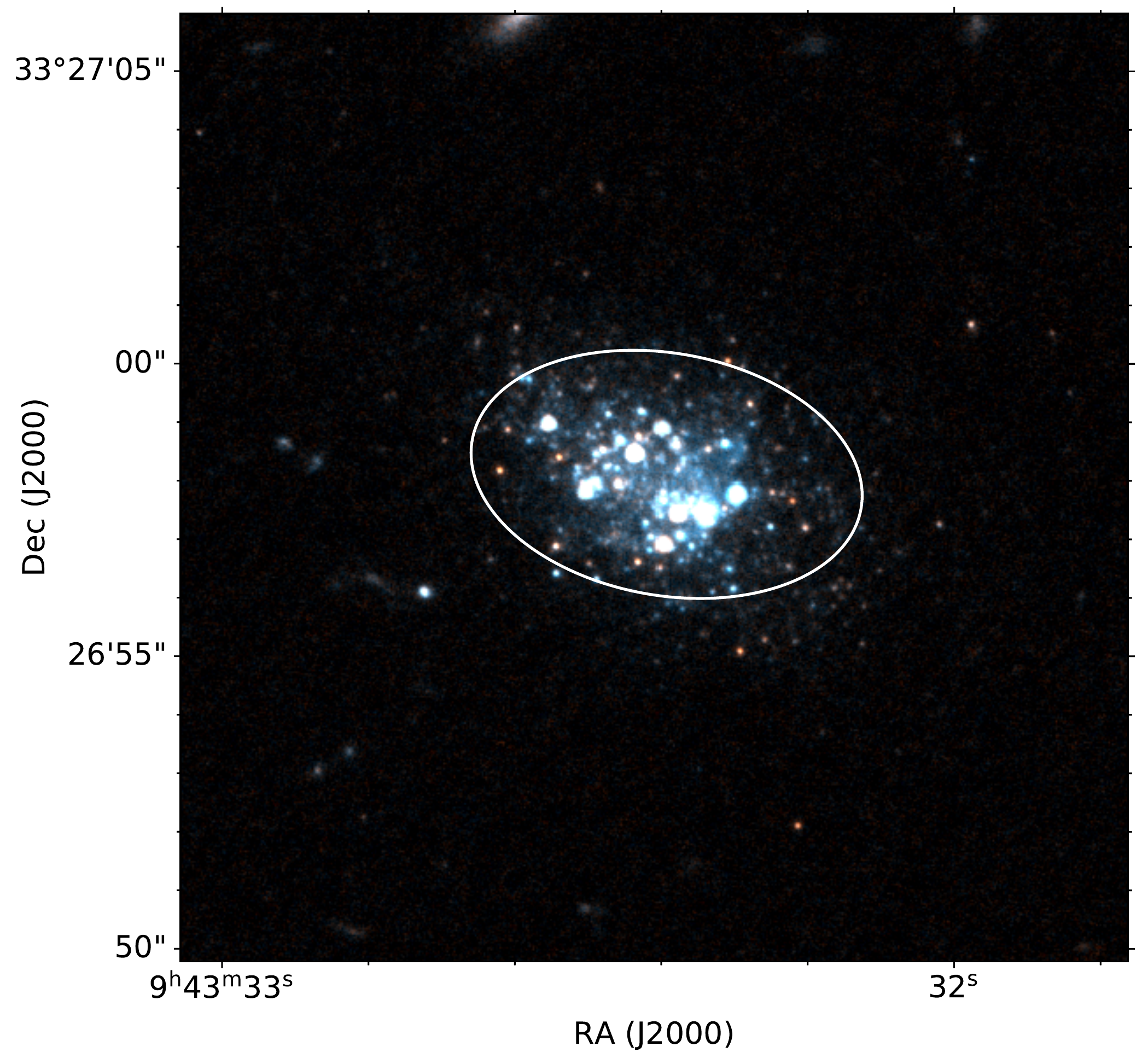}
\includegraphics[width=0.49\textwidth]{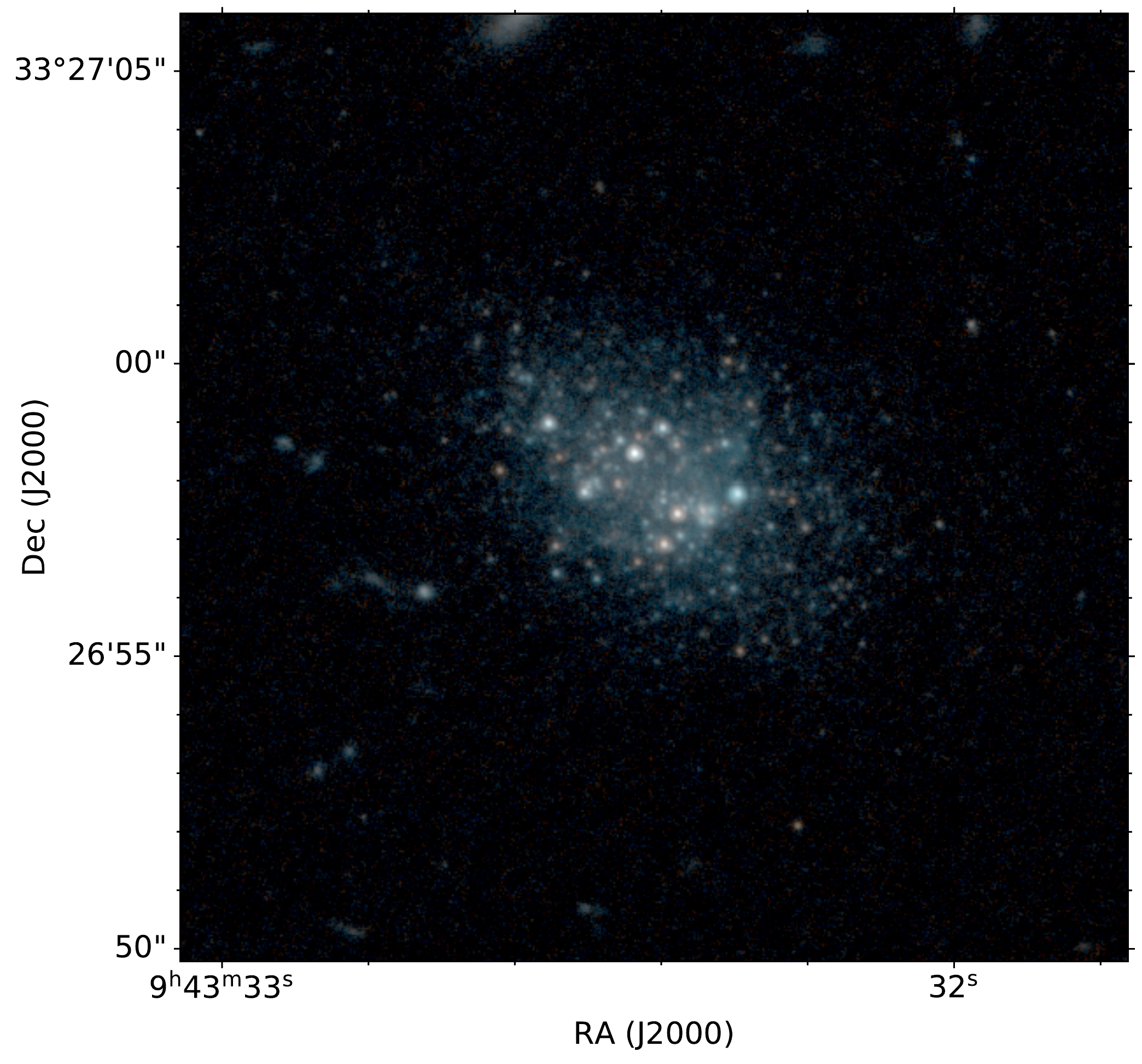}
\caption{Left: $HST$ WFC3 3-color image of Leoncino encompassing a field of view that is twice the optical diameter with North up and East left. The galaxy is compact, with blue stars dominating the center region, and has an extended, faint red population. The point sources within the ellipse are used in the CMD analysis below. Right: The same field of view shown with a log stretch that highlights the extended, lower luminosity, stellar distribution. }
\label{fig:image}
\end{figure*}

\section{HST Observations and Data Processing}\label{sec:obs}
\subsection{Observations}
New $HST$ observations of Leoncino were obtained using the Wide Field Camera 3 (WFC3) in the F606W and F814W filters between 2018 April 24 and 25 (HST GO-15243, PI McQuinn). The initial $HST$ ACS imaging from the SHIELD survey showed the galaxy to be highly compact with a crowded inner field. Thus, to increase resolution and reduce source blending, the WFC3 camera was chosen over the ACS instrument as WFC3 has 20\% smaller pixels. A total of 12 $HST$ orbits were split between the two filters with total integration times of 15.0 ks in the F606W filter and 18.6 ks in the F814W filter. The observations used the \textsc{wfc3-uvis-dither-line} small dither pattern of 2.5 pixels shifts in x and y between orbits to remove hot pixels and smooth the detector response. The galaxy was placed on the UVIS2 chip in quadrant C to minimize geometric distortions.

The images were processed by the standard WFC3 pipeline and corrected for charge transfer efficiency (CTE) nonlinearities caused by space radiation damage to the WFC3 instrument. The individual images (i.e., {\sc flc.fits} files) for each filter were median combined using the $HST$ Drizzlepac v2.0 software \citep{Gonzaga2012}, including cosmic-ray cleaning with {\sc Astrodrizzle}, and alignment with the task {\sc Tweakreg}.

\begin{figure*}
\includegraphics[width=0.32\textwidth]{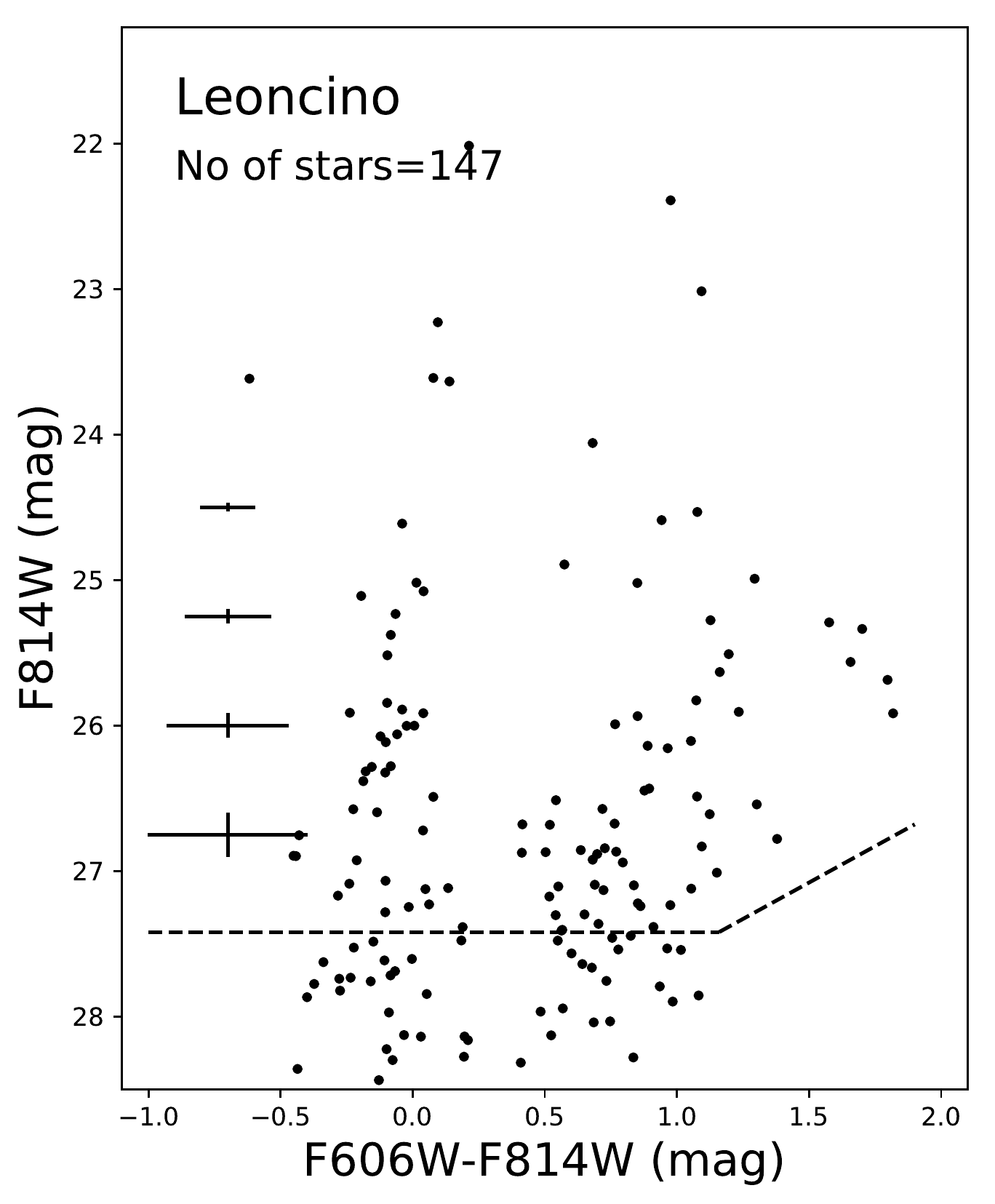}
\includegraphics[width=0.32\textwidth]{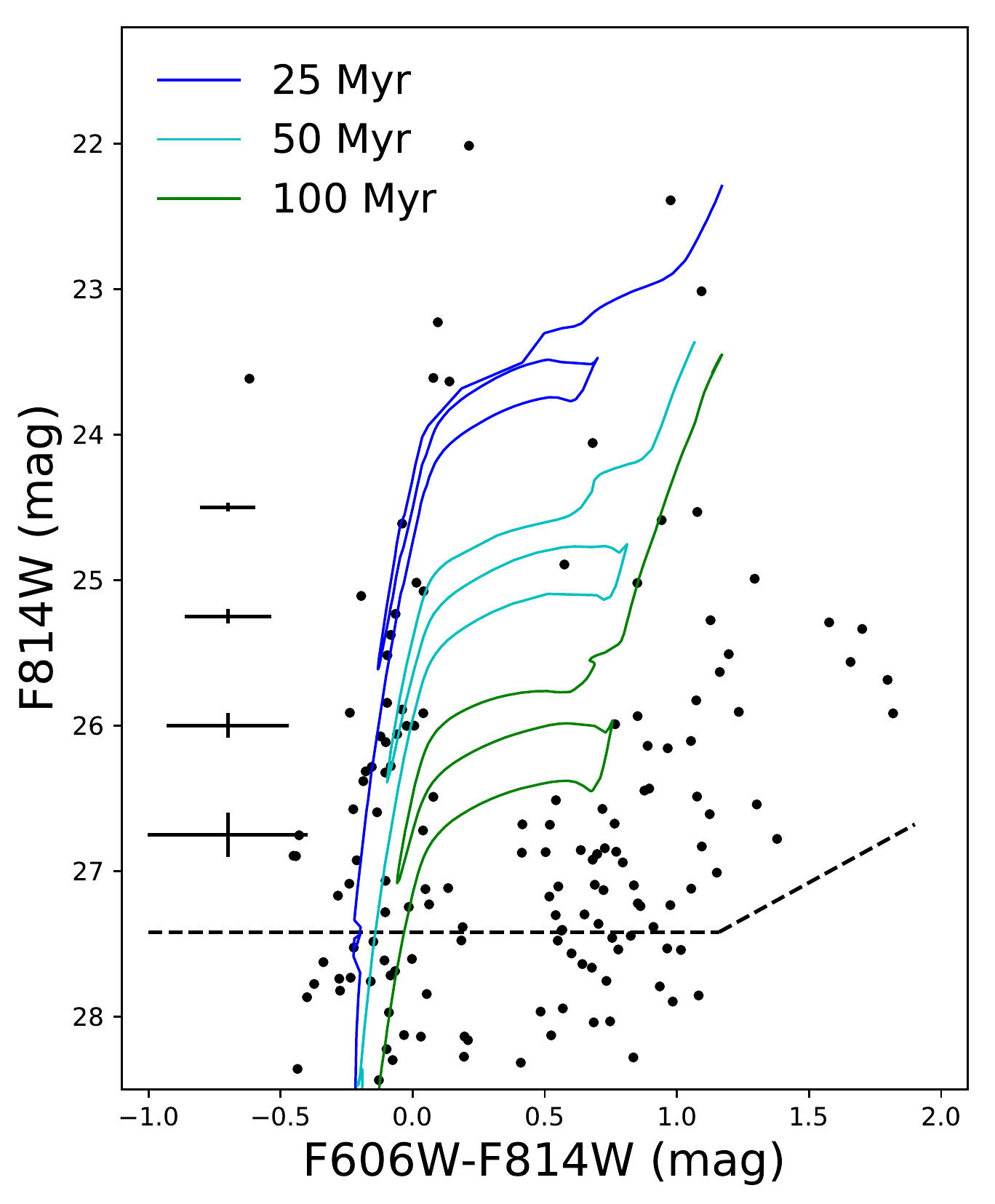}
\includegraphics[width=0.32\textwidth]{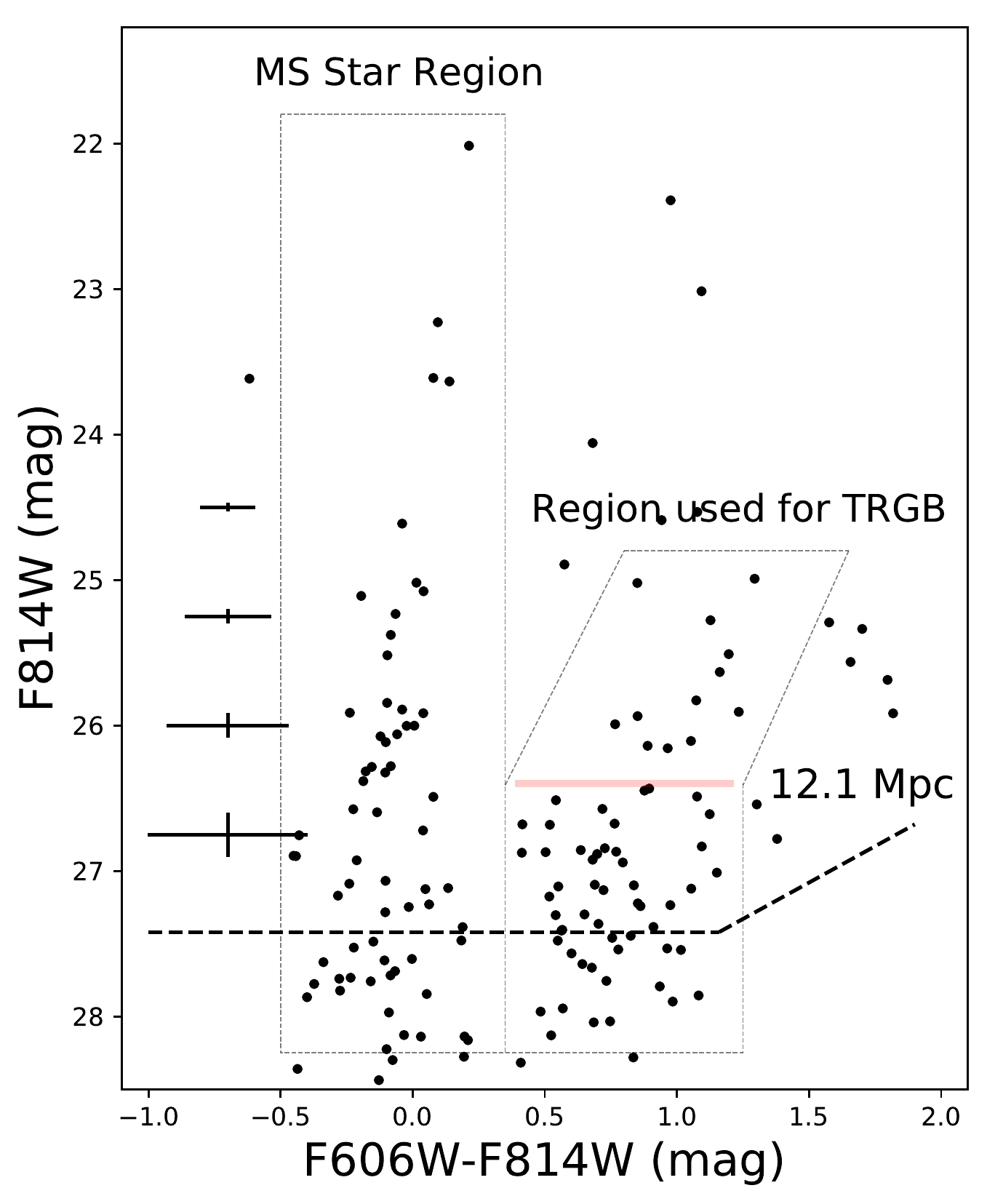}
\caption{Left: CMD of the 147 point sources in Leoncino within the ellipse shown in the left panel of Figure~\ref{fig:image} after applying quality cuts (see text for details) and correcting for Galactic extinction. There is a clear main sequence of stars with a population of redder sources that are some combination of RGB, red HeB, and AGB stars. Uncertainties are based on photometric uncertainties and incompleteness measured from artificial star tests. The dashed line is the 50\% completeness limit. Middle: CMD with PARSEC [M/H]$ = -2$ isochrones overlaid with ages of 25, 50, and 100 Myr. Right: CMD with the regions encompassing upper main sequence (MS) and, possibly, blue helium burning (BHeB) stars, and that used for the TRGB analysis outlined. The discontinuity in the F814W luminosity function, corresponding to a distance of 12.1 Mpc, is marked with a red line.}
\label{fig:cmd}
\end{figure*}

We present a color image of Leoncino from the $HST$ WFC3 imaging in the left panel of Figure~\ref{fig:image} with a linear stretch to highlight the data quality. The compact center of the galaxy is dominated by young stars and is surrounded by a larger stellar disk of fainter sources. A second image of the galaxy is shown in the right panel of Figure~\ref{fig:image} with a log stretch that highlights the lower luminosity outer extent of the stellar distribution. Both images encompass a field of view that is twice the optical diameter (see Table~\ref{tab:properties}) and were created using the drizzled image of the F814W data (red), an average of the F814W and F606W data (green) and the drizzled image of the F606W data (blue). The F814W and F606W drizzled images were created with the {\sc astrodrizzle} parameters {\sc final\_scale} set to 0.03 and {\sc final\_pixfrac} set to 0.6, making use of the sub-pixel dithering to improve the native pixel scale of the WFC3 instrument. A second combined image in the F606W filter was created at the native resolution and used as a reference image in the photometry, as described below. 

\subsection{Photometry}
Point-Spread Function (PSF) photometry was performed on the pipeline-processed, CTE-corrected, {\sc flc.fits} images using the WFC3 module of the {\sc dolphot} photometry package \citep{Dolphin2000}. The combined F606W images at the native WFC3 resolution were used as a reference image for identifying point sources and aligning the individual {\sc flc.fits} images. We used the {\sc dolphot} parameters adopted by the Panchromatic Hubble Andromeda Treasury survey \citep[PHAT;][]{Dalcanton2012}, because they are optimized for photometry in crowded fields. 

The photometry was filtered for well-recovered stars using standard quality cuts, including accepting sources with an error flag $<4$ and signal-to-noise ratios $\geq 4$ in both filters. In addition, only point sources with low sharpness and crowding values were included in the final photometry ([$F606W_{sharp} + F814W_{sharp}]^2 \leq 0.075$ and [$F606W_{crowd} + F814W_{crowd}] \leq 1.0$ mag). The sharpness parameter measures whether a point source is too sharp (similar to a cosmic ray) or too broad (indicating a background galaxy). The crowding parameter measures how much brighter a point source would be if nearby stars had not been fit simultaneously. The photometry was culled to include only stars within an ellipse centered on Leoncino. The spatial extent of the ellipse was empirically determined based on where the point source density dropped to the level of sources detected in an off-galaxy region of the image. The ellipse shown in Figure~\ref{fig:image} has a position angle of 80$^{\circ}$, eccentricity of 0.61, and semi-major axis of 6.\arcsec75 (see Table~\ref{tab:properties}). The final photometry was corrected for a modest amount of Galactic extinction ($A_{F606W} = 0.034$ mag; $A_{F814W} = 0.021$ mag) based on the dust maps of \citet{Schlegel1998} with a recalibration from \citet{Schlafly2011}. 

Artificial star tests were performed on the images to measure the photometric completeness using the same photometry package and filtered on the same quality cuts and spatial constraints as the observed stars. Approximately 500,000 artificial stars were injected in each of the 24 {\sc flc.fits} images following the spatial distribution of the photometry in the region of the galaxy. The 50\% completeness limits measured from these artificial star tests are F606W $=$ 28.58 mag and F814W $=$ 27.42 mag. The 50\% completeness limit is reached at a magnitude brighter than expected given the exposure times due to the compact nature of the galaxy and significant crowding in the galaxy's center.

The final color-magnitude diagram (CMD) for Leoncino, shown in the left panel of Figure~\ref{fig:cmd}, includes a total of 147 well-recovered stars. There is a clear sequence of blue stars containing upper main sequence and, possibly, BHeB stars, indicative of recent star formation. Representative uncertainties per magnitude are shown and include uncertainties from both the photometry and the artificial star tests. In the middle panel, we overlay stellar evolution isochrones from the PARSEC models \citep{Bressan2012} for stellar ages of 25, 50, and 100 Myr assuming a metallicity of [M/H]$ =-2$. The blue sequence of young stars in Leoncino, which is also highlighted in the right panel of Figure~\ref{fig:cmd}, is consistent with isochrone ages of $<100$ Myr.

The cadence of the 12 observing epochs, within the 23 hour total duration of the observations, was suitable for identifying short-period variable stars, but none were identified. Note that RR~Lyrae variable stars are far too faint to be detected in these observations at our derived distance. Cepheid variables would be possible to detect, but the small number of well-recovered stars makes detection statistically unlikely (and the $\sim$ one day duration of the observations means that only a fraction of the period of a bright Cepheid would be recorded). 

\section{The Distance to Leoncino}\label{sec:distance}
The WFC3 observations were designed to use the  tip of the red giant branch (TRGB) in the CMD as a distance indicator. The TRGB distance method is a standard candle approach based on calibrating the sharp discontinuity in a CMD at the upper boundary of the RGB stars, just prior to the helium flash, to an absolute luminosity. Robustly measuring the TRGB requires a sufficient number of stars in the CMD to both unambiguously identify the RGB sequence and to fully populate the TRGB. Previous studies of low-mass systems have shown that a minimum of $\sim300$ stars in the magnitude below the TRGB will result in robustly measured distances with uncertainties of order 0.1 mag \citep{Makarov2006, McQuinn2013}. 

\begin{figure}
\includegraphics[width=0.48\textwidth]{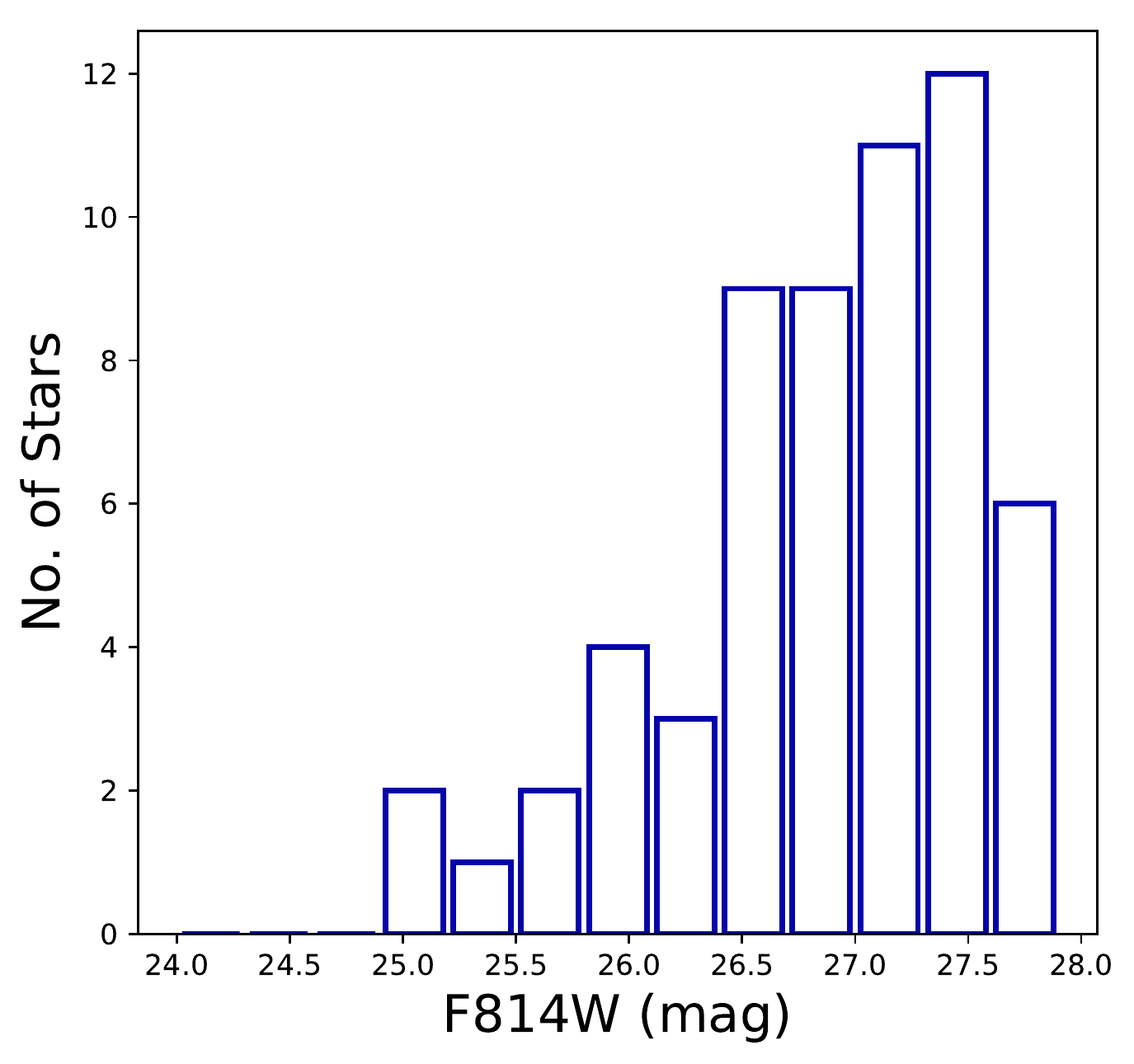}
\caption{Extinction corrected F814W luminosity function from point sources in the RGB region of the CMD (marked with a polygon in the right panel of Figure~\ref{fig:cmd}). There is a  discontinuity of the luminosity function at 26.4 mag, which we interpret as the maximum luminosity of RGB stars just prior to the helium flash. We adopt this as the magnitude of the TRGB.}
\label{fig:lf}
\end{figure}

\begin{figure*}
\includegraphics[width=0.98\textwidth]{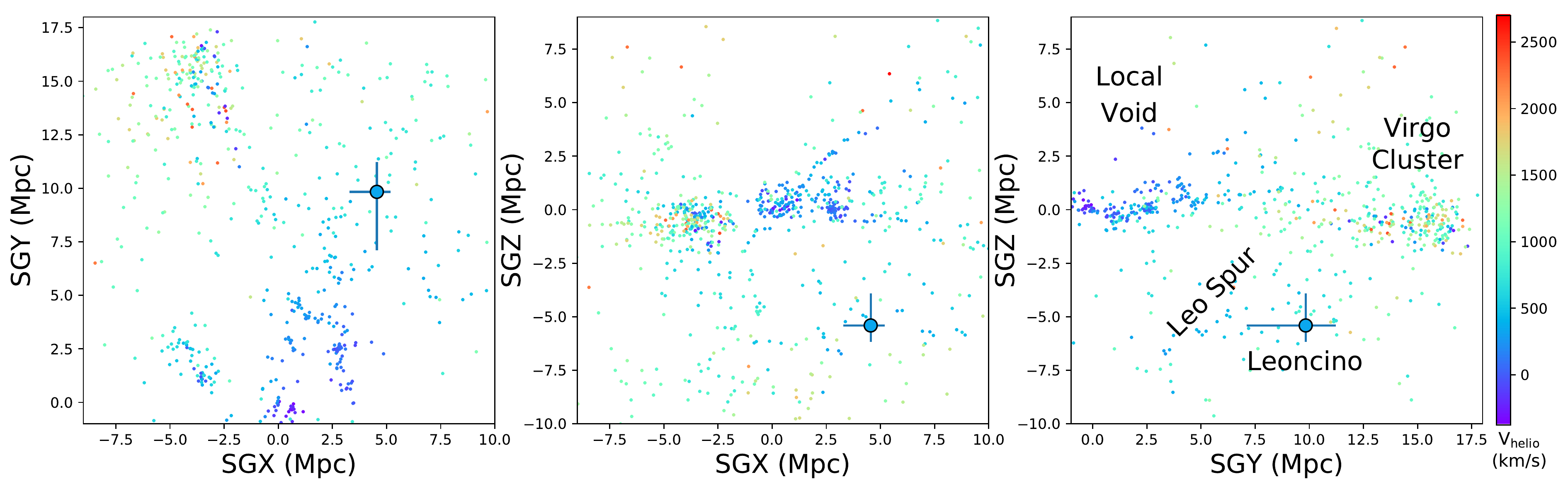}
\caption{Sources within 18 Mpc from the Cosmicflows-3 database plotted in projected supergalactic coordinates. The points are color-coded by velocity; see side color bar for values. The location of Leoncino is marked with a blue circle, corresponding to its V$_{\rm helio}$ of 514 km s$^{-1}$. Major gravitational structures are labeled in the last panel and include the Local Void with an approximate centroid near SGX, SGY, SGZ coordinates (0,0,5), the Virgo Cluster ($-$5,16, 0), and the Leo Spur (5,5,$-$5). Leoncino lies in an under-dense region, catalogued as void number 12 in \citet{Pustilnik2019}.}
\vspace{20pt}
\label{fig:environment}
\end{figure*}

The CMD shown in Figure~\ref{fig:cmd} is sparsely populated, with 147 high-fidelity point sources measured in the galaxy from observations reaching F814W $\sim28$ mag. The low star counts are what is expected of a galaxy with stellar masses of a few $\times10^5$ \msun\ based on modelling synthetic CMDs with stellar evolution libraries \citep{McQuinn2013}. The sparse nature of the CMD makes it challenging to unambiguously identify the TRGB with high precision. 

Because the TRGB technique relies on identifying the break in the luminosity function of the RGB stars, we avoid including stars that are clearly not in the RGB region of the CMD in our analysis. The CMD in Figure~\ref{fig:cmd} has a distinct sequence of blue stars corresponding to upper main sequence (MS) and, possibly, blue helium burning (BHeB) stars (labeled MS star region; marked with a rectangular outline) and a sequence of red stars (F606W$-$F814W $\gtsimeq 0.35$ mag) that is a combination of red helium burning (RHeB) stars, asymptotic giant branch (AGB) stars, and RGB stars. While the most luminous of these red stars are likely RHeB stars, the fainter stars are RHeB, AGB,  or RGB stars. Assuming the average stellar metallicity in Leoncino is approximately equal to the gas-phase metallicity, the RGB is expected to be at an F606W $-$ F814W color of $\sim1$ based on PARSEC stellar evolution isochrones \citep{Bressan2012}. Using the distribution of red point sources in the CMD and the expected color of the RGB as guides, we select point sources from this red sequence for TRGB analysis; this region is marked with a polygon in Figure~\ref{fig:cmd}. The bright sources that are redder than this region (i.e., F606W$-$F814W $\gtsimeq 1.5$) are likely AGB stars. 

Given the small number of point sources, we use a simple analysis to identify the TRGB. Figure~\ref{fig:lf} presents the extinction corrected F814W luminosity function of stars inside the region marked for TRGB analysis in Figure~\ref{fig:cmd}, with bin sizes of 0.3 mag. A strong discontinuity in the F814W luminosity function occurs in the histogram bin edge at 26.4 mag, with a drop from 9 stars to 3 stars in the adjacent bin. Assuming Poissonian fluctuations for an uncertainty, a jump from 9 $\pm$3 to 3$\pm1.7$ is statistically significant at the 1.7$\sigma$ level. There are other, smaller drops in the LF (for example, at 25.9 mag), but these are below the 1$\sigma$ significance level. The discontinuity identified in the luminosity function maps to the change in point source density at F814W$=26.4$ mag in the CMD, marked with a red line in Figure~\ref{fig:cmd}. The populations of red stars above this magnitude are candidate RHeB stars and AGB stars. 

A TRGB mag of 26.4 mag corresponds to a distance modulus of 30.4 mag, using the zero-point calibration of the TRGB specific to the HST WFC3 filters from \citet[][their averaged blue $I$ calibration]{Jang2017}. For the lower measurement uncertainty, we adopt 0.30 mag, based on the width of the luminosity function histogram. For the upper measurement uncertainty, we adopt a more conservative estimate. It is possible that the TRGB lies above our identified discontinuity at 26.4 mag, but belies detection due to low star counts. As mentioned above, synthetic CMDs generated using stellar evolution libraries show that the upper RGB will be under-populated at very low galaxy masses. Thus, our detected discontinuity may correspond to RGB stars \textit{fainter than} the actual TRGB (i.e., corresponding to a {\it farther} distance). From the CMD in Figure~\ref{fig:cmd}, there are 66 stars in the region used for the TRGB analysis, which is approximately equal to the 69 stars found in a similarly sized RGB region in the synthetic CMD of a $3\times10^5$ \msun\ galaxy in \citet[][their Appendix]{McQuinn2013}. Leoncino also has a loose group of seven stars brighter than the discontinuity in the luminosity function, similar to the 8 stars in the modelled CMD. The TRGB in the synthetic CMD lies 0.6 mag above the identified discontinuity. We conservatively adopt 0.6 mag as the upper measurement uncertainty on the distance modulus for Leoncino, corresponding to a range of closer distances. This follows the same approach used for the similarly low-mass galaxy Leo~P, whose initial TRGB distance measurement yielded 1.72$^{+0.14}_{-0.40}$ Mpc with lower uncertainties based on the same modelling \citep{McQuinn2013}, and whose final distance measurement from horizontal branch and RR~Lyrae stars was closer at 1.62$\pm0.15$ Mpc \citep{McQuinn2015e}. Summing our uncertainties for Leoncino in quadrature with the zero-point calibration uncertainties, the final distance modulus is 30.4$^{+0.31}_{-0.60}$ mag, or a distance of 12.1$^{+1.7}_{-3.4}$ Mpc.

Table~\ref{tab:properties} provides a list of distance-dependent values. Based on the adopted distance of 12.1 Mpc, the $M_B$ luminosity of Leoncino is $-10.58^{+0.28}_{-0.07}$ mag. The physical size of the galaxy is small, with a major axis diameter of only 0.8 kpc. We estimate the stellar mass in Leoncino to be M$_* = (7.3^{+2.2}_{-4.3})\times10^5$ \msun, based on the integrated flux from \textit{Spitzer Space Telescope} IRAC 3.6~$\micron$ imaging of Leoncino (J. M. Cannon et al., in preparation), and assuming a mass-to-light ratio of 0.47 \citep{McGaugh2014}. This value is in reasonable agreement with the stellar mass of 3$\times10^5$ \msun\ in the synthetic galaxy from \citet{McQuinn2013} used as a basis to estimate the upper uncertainty on the distance. 

We use these properties to compare Leoncino with other extremely low-metallicity galaxies and provide context for the possible chemical evolution pathways of the galaxy. 

\section{The Environment Surrounding Leoncino}\label{sec:env}
The distance determination of 12.1$^{+1.7}_{-3.4}$ Mpc allows us to explore the 3-dimensional environment around the Leoncino dwarf galaxy. We made use of the Cosmicflows-3 extragalactic database \citep[CF3;][]{Tully2016} which tabulates the distances to 18,000 galaxies in the nearby universe. We selected all galaxies with distances less than 18 Mpc and transformed their sky coordinates to the supergalactic cartesian coordinate system (SGX, SGY, SGZ):

\begin{subequations}
\begin{align}
{\rm SGX = Distance \cdot cos(SGL) \cdot cos(SGB)} \\
{\rm SGY = Distance \cdot sin(SGL) \cdot cos(SGB)} \\
{\rm SGZ = Distance \cdot sin(SGB)}
\end{align}
\label{eq:sg_coordinates}
\end{subequations}

\noindent where SGL and SGB are the supergalactic longitude and latitude respectively. In supergalactic coordinates, the Milky Way lies at SGX, SGY $=(0,0)$ and the supergalactic plane aligns with the SGZ $=0$ plane. 

\begin{figure*}
\includegraphics[width=0.98\textwidth]{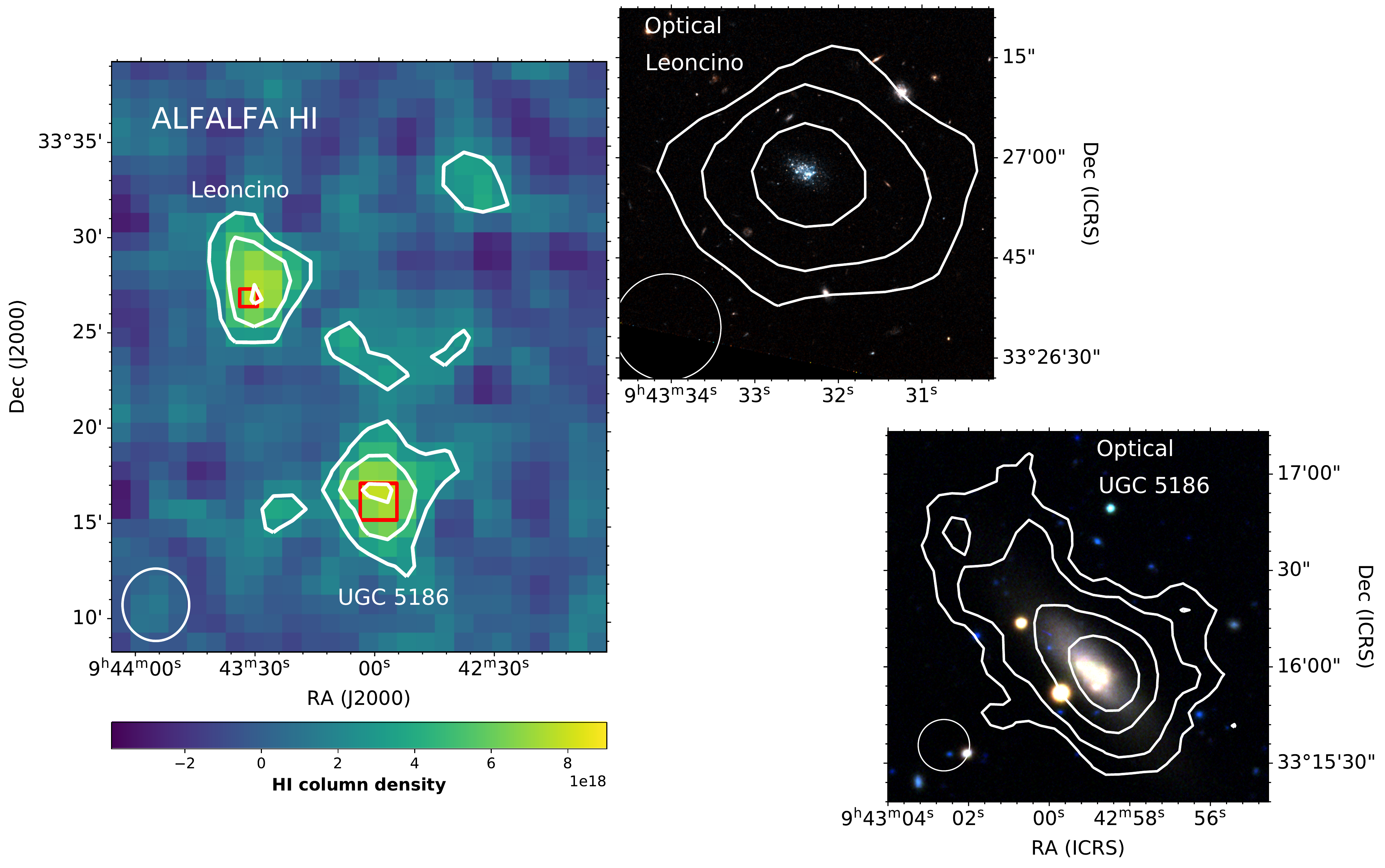}
\caption{Left: ALFALFA \hi\ map of the region including Leoncino and the nearby galaxy UGC~5186, plotted to a column density of 10$^{18}$ cm$^{-2}$. \hi\ contours correspond to \hi\ column densities of (2.5, 5, 7.5) $\times10^{18}$ cm$^{-2}$ and the \hi\ beam size is shown in the lower left. There is a putative detection of \hi\ between the two galaxies. Top Right: HST image of Leoncino with \hi\ contours from the JVLA overlaid. The galaxy is extremely gas-rich with the \hi\ extending significantly farther than the stellar distribution. Bottom Right: Optical image of UGC~5186 from the Dark Energy Camera Legacy Survey (DECaLS) with \hi\ contours from the JVLA overlaid. The JVLA \hi\ contours correspond to \hi\ column densities of (1, 2, 4, 8) $\times10^{20}$ cm$^{-2}$. No \hi\ was detected between the galaxies in the JVLA data. The zoomed-in optical images of Leoncino and UGC~5186 are approximately aligned with their orientation on the sky, highlighting the outer \hi\ morphology of UGC~5286 is elongated in the projected direction of Leoncino.}
\label{fig:hi}
\end{figure*}

Figure~\ref{fig:environment} shows the distribution of the CF3 data color-coded by the heliocentric velocities of the galaxies. The large, light blue point marks the location of Leoncino in supergalactic coordinates, with uncertainties. The major gravitational structures in the nearby universe are readily apparent and marked in the third panel. The Local Sheet of galaxies $-$ which includes our Milky Way $-$ is located at SGZ $\sim$ 0. Just above the Local Sheet at positive SGZ values is the Local Void, expanding towards us \citep{Tully2008}. The Virgo Cluster includes the largest concentration of galaxies and lies roughly along the supergalactic plane at a distance of $\sim16.5$ Mpc in the SGY direction \citep{Mei2007}. 

Below the supergalactic plane lies an additional structure that is thought to be gravitationally linked, namely the Leo Spur \citep{Tully2008, Karachentsev2015}, which stretches roughly from SGY, SGZ $=(3,-6)$ to $(10,0)$. Several galaxies in the Leo Spur have robustly measured TRGB distances ranging from $\sim8-12$ Mpc \citep{McQuinn2014, Karachentsev2015} with radial velocities less than 500 km s$^{-1}$, placing the structure in the foreground of the Virgo Cluster. The lower uncertainties on the distance to Leoncino overlap with the edge of the Leo Spur and the velocity of Leoncino is consistent with the velocities measured for Leo Spur galaxies. It is possible that Leoncino lies at the edge of this structure. 

Overall, Leoncino is located in an under-dense region or void, well inside an area corresponding to Void No.~12 in the void catalog of \citet{Pustilnik2019}. We found no neighboring galaxies within 1 Mpc of Leoncino in the CF3 data set nor in the Updated Nearby Galaxy Catalog \citep{Karachentsev2013}. Note that this does not preclude the possibility that one or more of these galaxies are in the vicinity of Leoncino. The distances to the majority of the sources in the full CF3 database (67\%) were determined using the Tully-Fisher (TF) relation, which have a minimum uncertainty of $\sim$20\% \citep[e.g.,][]{Tully2013}. At the nominal distance to Leoncino, this translates to a distance uncertainty of 2.4 Mpc.  Note also, that in specific cases, TF distances can be uncertain by a factor of two or more \citep[see, e.g.,][for comparisons of TF distances with more secure distance methods]{McQuinn2016a, McQuinn2016b, McQuinn2017b}.

We did identify two galaxies that are nearby on the sky and with similar radial velocities to Leoncino using the ALFALFA \hi\ data set. The first is UGC~5209 (KKH~54) located 75\arcmin\ from Leoncino in projection, which translates to a minimum separation of 264 kpc at the distance of Leoncino. UGC~5209 has a heliocentric velocity, V$_{\rm helio}$ of 538 km s$^{-1}$, which is comparable to V$_{\rm helio}$ of 514 km s$^{-1}$ of Leoncino. UGC~5209 has a TRGB distance of 10.42$\pm0.35$ Mpc \citep{Karachentsev2015} and is located in the same Void No.~12 as Leoncino \citep{Pustilnik2019}. 

The second is UGC~5186, located 13\arcmin\ in projection from Leoncino, which translates to a minimum separation of 46 kpc at the distance of Leoncino, with a heliocentric velocity V$_{\rm helio}$ of  549 km s$^{-1}$. 

UGC~5186 is also found in Void No.~12 \citep{Pustilnik2019}. The galaxy has a distance estimate of 10.8 Mpc based on the velocity with a correction for local motion \citep{Pustilnik2011a} and an alternate distance of 8.3 Mpc based on the TF method \citep{Tully2013}. The flow-model distance is favored in \citet{Pustilnik2011a} based on studying the motions of galaxies in Void No.~12, which we adopt at the best estimate of the distance. 

The ALFALFA \hi\ map, shown in the left panel of Figure~\ref{fig:hi}, suggests there is \hi\ gas connecting Leoncino and UGC~5186. Follow-up, higher spatial resolution JVLA data show the \hi\ distribution in UGC~5186 is elongated towards Leoncino, as seen in the lower right panel in Figure~\ref{fig:hi}, which also suggests an ongoing gravitational interaction between the two galaxies. No low surface brightness gas between the two galaxies is detected in the higher resolution JVLA data, but the column density detection limit is a factor of 10 higher than the ALFALFA data. Despite the physical separation implied by the distances of 12.1 Mpc to Leoncino and the flow-model distance of 10.8 Mpc to UGC~5186, the uncertainties on both distances, similar radial velocities, and \hi\ distributions leaves open the possibility that UGC~5186 is a close neighbor to Leoncino and the two are interacting. We discuss the impact that such an interaction could have in triggering the recent star formation activity in Leoncino in \S\ref{sec:discuss}. 

\begin{figure}
\includegraphics[width=0.48\textwidth]{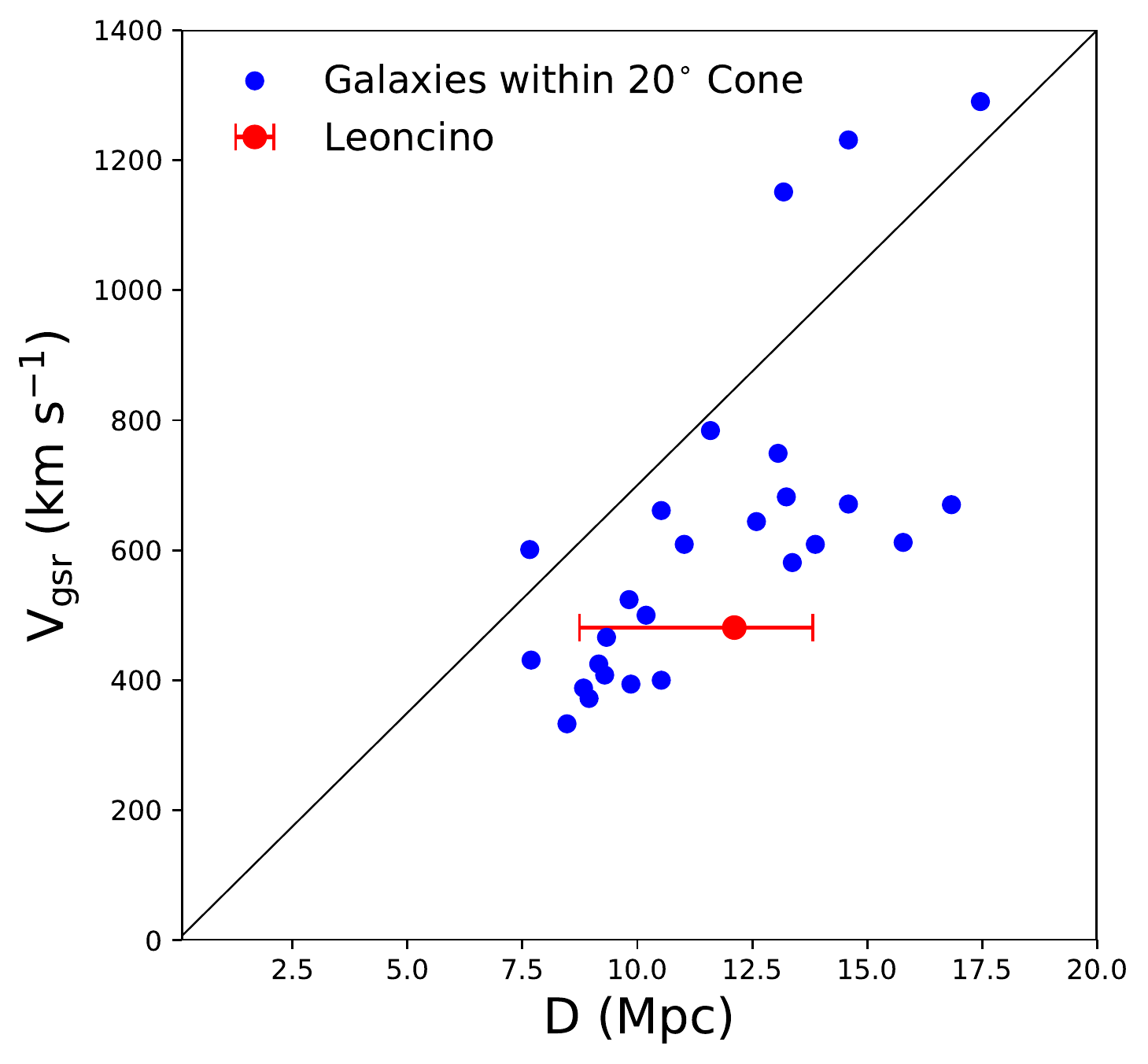}
\caption{ Galactic standard of rest velocities (V$_{\rm gsr}$) versus distance for galaxies within a 20 degree cone of Leoncino from the Cosmicflows-3 database \citep{Tully2016}. The solid line represents a Hubble-Lema{\^i}tre law assuming a Hubble constant of 70  km s$^{-1}$ Mpc$^{-1}$. The majority of galaxies in this region of the sky are known to have peculiar velocities and lie below this line. At the adopted distance of 12.1 Mpc, Leoncino is consistent with other galaxies nearby on the sky.}
\label{fig:hubble_diagram}
\end{figure}

The Hubble-Lema{\^i}tre diagram in Figure~\ref{fig:hubble_diagram} allows us to compare the velocity, measured in the Galactic standard of rest, V$_{\rm gsr}$, and the distance to Leoncino with other galaxies in the nearby universe. We selected all galaxies from the CF3 extragalactic database \citep{Tully2016} with distances less than 20 Mpc and within a 20$^{\circ}$ cone of Leoncino, corresponding to a cone with a physical diameter of $\sim4$ Mpc at the distance of Leoncino. Because of the gravitational pull of the Virgo cluster, this volume in space is known to have peculiar flow velocities. As a result, many systems are offset from a Hubble-Lema{\^i}tre law expectation assuming an approximate Hubble constant value of $\sim70$ km s$^{-1}$ Mpc$^{-1}$. Based on our measured distance, Leoncino is consistent with the Hubble flow in this region, as shown in Figure~\ref{fig:hubble_diagram}.

\vspace{40pt}
\section{Leoncino on the LZ and MZ Relations}\label{sec:lz}
An important aim of the present work is to explore the chemical evolution pathway of the XMP galaxy Leoncino. Thus, in this section, we examine the location of Leoncino in the LZ and MZ planes relative to other star-forming low-mass galaxies. We start with the LZ relation and present the oxygen abundances as a function of absolute luminosity, as the luminosity is the more readily available measurement from different studies. We then examine the location of Leoncino in the MZ relation. While stellar masses are not available for as many galaxies in the comparison samples, the MZ relation traces the more fundamental quantity of stellar mass and is less impacted by fluctuations in recent star-formation rates than the LZ relation. 

\subsection{Oxygen Abundances}
For all examinations of the LZ and MZ relations, we restrict the comparisons to galaxies with oxygen abundances determined using the direct method from the temperature sensitive [O III] $\lambda4363$ auroral line. Other theoretical and empirical abundance methods that depend solely on the ratio of strong emission lines have been shown to vary by different amounts depending on the methods and models used, as well as the properties of the calibration samples \citep[e.g.,][]{vanZee2006, Kewley2008, Moustakas2010, Andrews2013}. Therefore, to conduct a uniform analysis, we do not include oxygen abundances derived using strong-line methods in our comparisons. 

\subsection{The Luminosity-Metallicity Relation}
Luminosity measurements are readily available for a larger sample of galaxies than the derived quantity of stellar mass. To place the properties of Leoncino in a broader context, we present a comparison of the oxygen abundances of low-mass galaxies as a function of absolute optical luminosity. We include galaxies in typical field environments in the Local Volume from \citet{Berg2012}, galaxies in void environments from \citet{Pustilnik2016} and \citet{Kniazev2018}, and XMP galaxies from various studies. 

\subsubsection{Local Volume Legacy Galaxies}
\begin{figure*}
\includegraphics[width=0.48\textwidth]{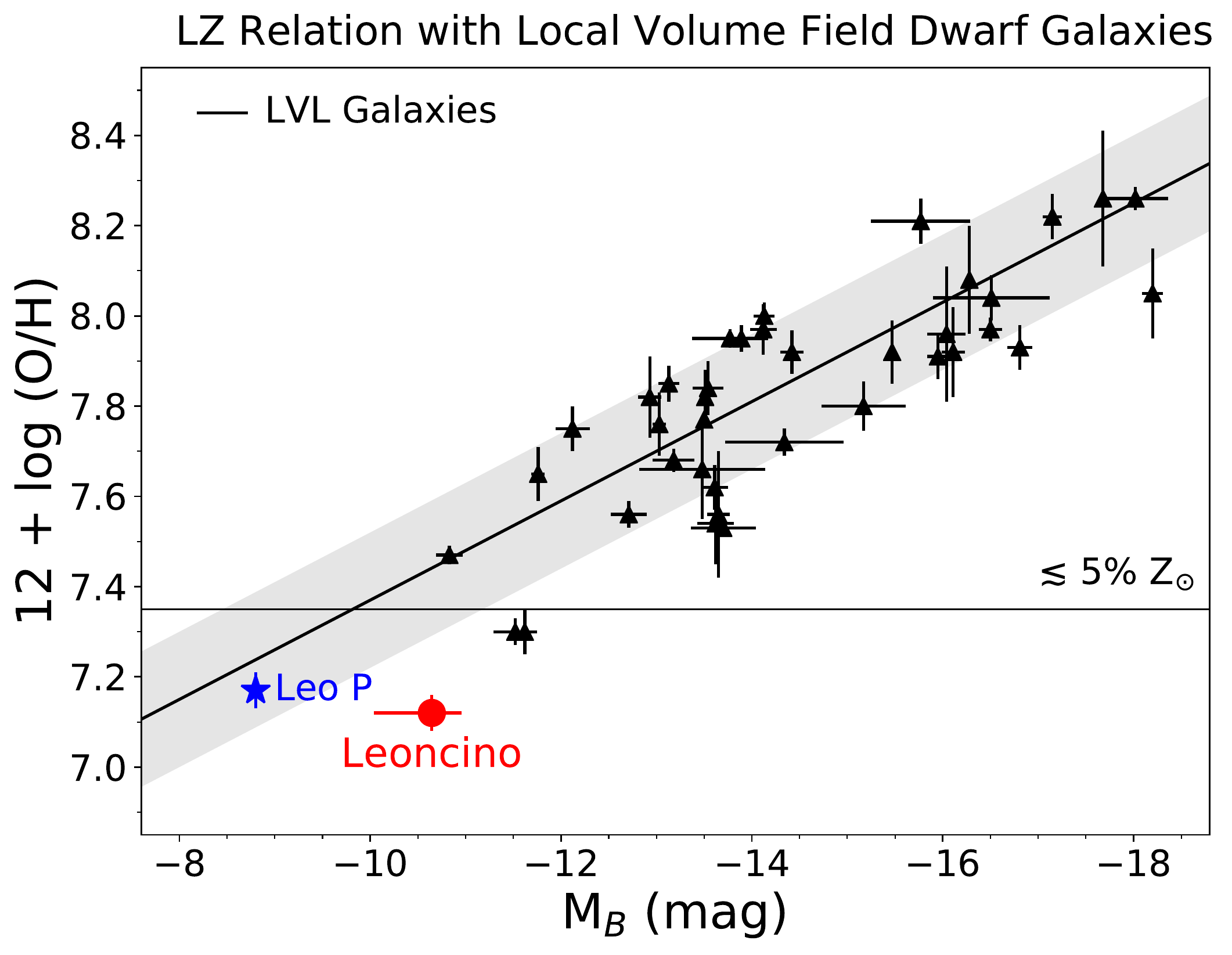}
\includegraphics[width=0.48\textwidth]{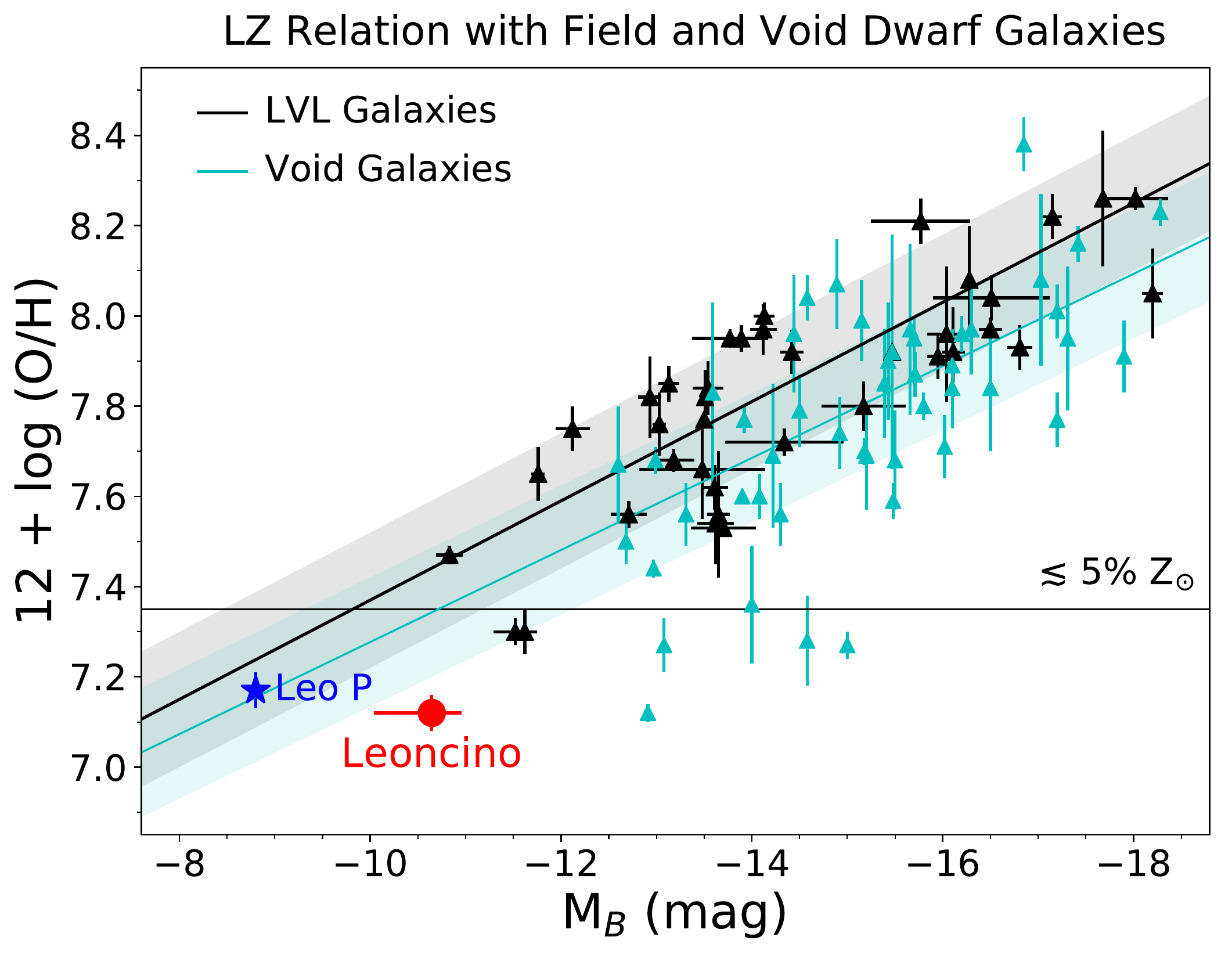}
\caption{Luminosity ($M_B$) vs. $12+$log(O/H). The luminosity of Leoncino is based on the adopted distance of 12.1 Mpc. Left: Typical, star-forming low-mass galaxies with direct-method oxygen abundances and robust distances from the Local Volume Legacy (LVL) sample form a tight relation \citep[][black triangles]{Berg2012}. The Leoncino dwarf, which hosts a population of young, luminous stars, lies slightly off the LZ relation. Leo~P, a comparable-mass galaxy that hosts fewer young stars, lies on the LZ relation. Right panel: We add galaxies in void environments that have direct-method oxygen abundances \citep[][small cyan triangles]{Pustilnik2016, Kniazev2018}. The best-fitting line and dispersion to the void galaxies is shown in cyan. When galaxies with only direct-method oxygen abundances are included, the LZ relation for void galaxies is consistent with the relation for galaxies in field environments.}
\label{fig:lz_berg_voids}
\end{figure*}

We begin with a comparison of the oxygen abundance as a function of luminosity for Leoncino and typical, low-mass, star-forming galaxies in the nearby universe, shown in the left panel of Figure~\ref{fig:lz_berg_voids}. The sample of low-mass galaxies is drawn from \citet{Berg2012} and includes the subset of galaxies in the Local Volume Legacy (LVL) survey with direct-method oxygen abundances and high-quality distances (i.e., their `Combined Select' sample). The LVL galaxies lie along a tight correlation with a best fit line of 6.27$\pm0.21 - (0.11\pm0.01) M_B$ and an intrinsic dispersion in log(O/H) of $\sigma = 0.13$, reproduced on the plot as a solid black line with the $1\sigma$ range shaded in grey.

Leoncino, shown in red in Figure~\ref{fig:lz_berg_voids}, lies somewhat off the LZ relation. Despite Leoncino's low stellar mass of 7.3$\times10^5$ \msun, the galaxy hosts a significant number of young, luminous, upper main sequence stars (seen in the CMD of Figure~\ref{fig:cmd}), which suggests that the offset from the LZ relation may be due to the higher luminosity of these stars. We examine this quantitatively in \S\ref{sec:pathways}. Also shown is the galaxy Leo~P, similarly discovered via its \hi\ content in the ALFALFA survey, which agrees with the LZ relation \citep{Skillman2013}. Leo~P has a comparable stellar mass of 5.6$\times10^5$ \msun, but significantly fewer stars in the upper main sequence region \citep{McQuinn2015e}. 

\subsubsection{Void Galaxies}
Next, as Leoncino is located in a void \citep{Pustilnik2019}, we consider a population of galaxies in low-density regions and where they lie relative to the LZ relation. Recent studies have compared the oxygen abundance of galaxies in void environments with galaxies in more typical field environments. Void galaxies are reported to lie in a parallel sequence to the \citet{Berg2012} relation for typical LVL field galaxies, but offset to lower oxygen abundances by 30-40\%, in a comparison of 81 low-mass systems in the Lynx-Cancer void \citep{Pustilnik2016} and 36 similar systems in the Eridanus void from the same project \citep{Kniazev2018}. Oxygen abundances were measured via the direct method when the [O III] $\lambda$4363 line was detected and via the strong emission lines using semi-empirical and empirical methods when it was undetected. 

Adopting the same criteria for our comparison of galaxies in void environments, we selected only the galaxies from the void survey program of \citet{Pustilnik2016} with direct-method abundances and overplot them as blue triangles in right panel of Figure~\ref{fig:lz_berg_voids}. The sample represents 31 galaxies of the Lynx-Cancer void sample from \citet{Pustilnik2016} and 16 galaxies from the Eridanus void sample from \citet{Kniazev2018}. Most of the selected galaxies are consistent with the LZ relationship defined by Local Volume galaxies by \citet{Berg2012}, with a small bias toward lower values of O/H. 

To compare the groups statistically, we performed a linear regression on the void galaxy data. We excluded the three XMP galaxies that are significant outliers from the rest of the sample in Figure~\ref{fig:lz_berg_voids}. Galaxies that lie away from the LZ relation are thought to have followed different evolutionary pathways, perhaps as a result of tidal interactions that lead to gas infall and more effective mixing in the ISM \citep[e.g.,][and see discussion below]{Ekta2010b}. The best-fitting line to this sub-sample with direct method measurements is 6.26$\pm0.25 - (0.10\pm0.02) M_B$ and an intrinsic dispersion in log(O/H) of $\sigma = 0.14$, shown as a cyan line with the 1$\sigma$ range shaded. This best-fitting line for the void galaxies lies below but is consistent with the relation for low-mass galaxies in typical field environments from \citet{Berg2012}. If we include the three XMP galaxies that are offset from the trend, the two relations are still consistent with one another but the best-fitting line to the void galaxies has a steeper slope than the one shown (5.91$\pm0.30 - (0.12\pm0.02) M_B$ and an intrinsic dispersion in log(O/H) of $\sigma = 0.19$). Having interferometric \hi\ observations of these three systems would be highly desirable in order to test the hypothesis that their low abundances are externally driven by interactions. 

In summary, when only direct-method abundances are considered, the best-fitting lines between the void dwarf galaxies and gas-rich dwarf galaxies in more typical field environments are separated by $<1\sigma$. Quantitatively, the dispersions from the two lines combined in quadrature yield a value of 0.21, which can be compared to the separation of 0.16 at $M_B = -18$ mag and 0.12 at $M_B = -13$ mag. The void galaxies are biased towards lower abundances, but the difference is quite small. Since our direct-method abundance criteria removed a significant number of void galaxies in the original sample, follow-up direct-method measurements are needed to confirm this result. 

\subsubsection{XMP Galaxies: 12$+$log(O/H) $\leq 7.35$}\label{sec:XMP}
\begin{figure*}
\includegraphics[width=0.98\textwidth]{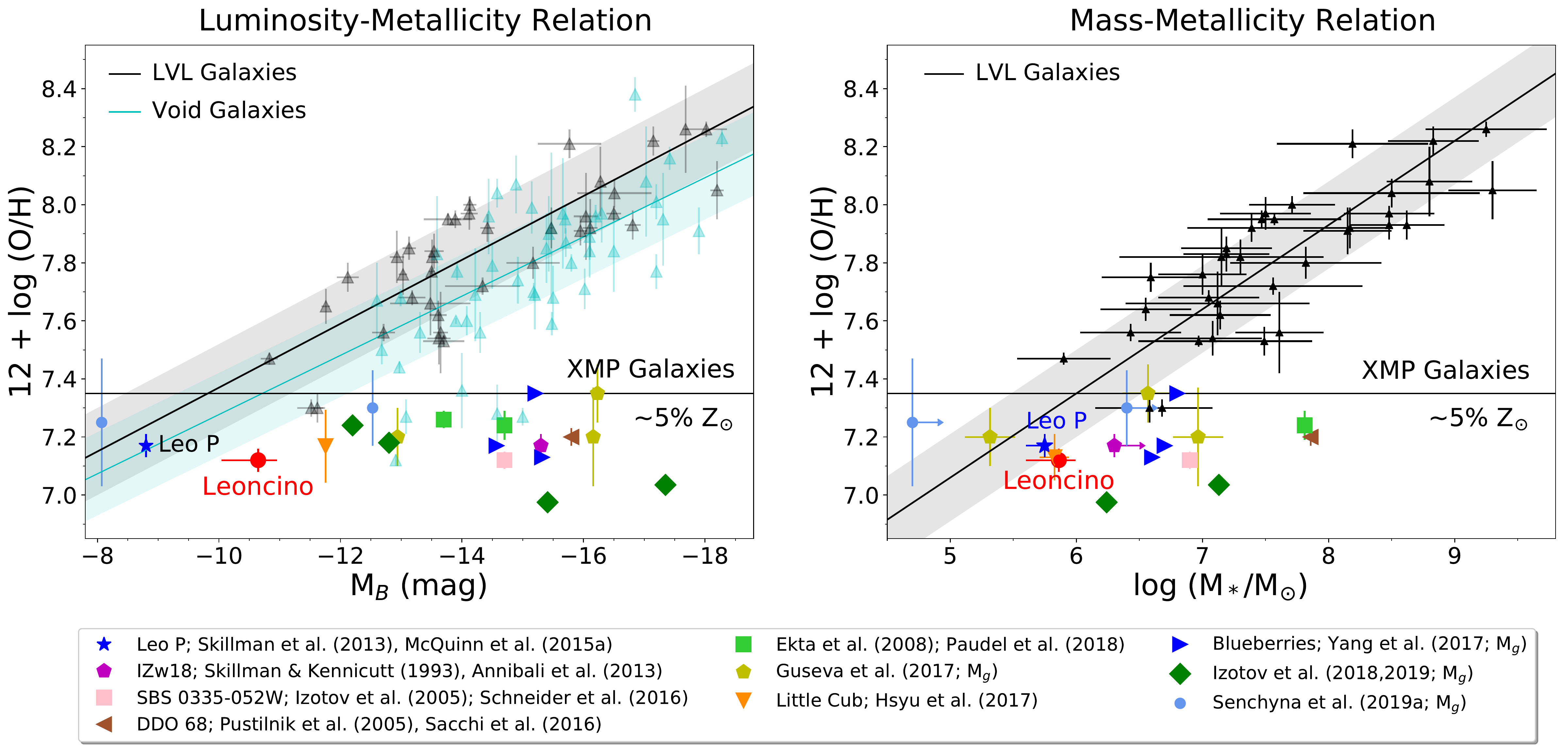}
\caption{Left panel: Luminosity vs.~$12+$log(O/H) reproduced from Figure~\ref{fig:lz_berg_voids} with the addition of other XMP galaxies with oxygen abundances below 5\% \zsun\ from various studies measured with the direct-method. All XMP galaxies are offset from the LZ relation, with the exception of the Leo~P galaxy whose low metallicity has been shown to driven by inefficient star formation and galactic winds \citep{McQuinn2015f} and J1005$+$3772 \citep{Senchyna2019a}. If the young, upper main sequence stars in Leoncino are excluded, Leoncino agrees with the LZ relation based on the estimated fainter luminosity. Uncertainties in oxygen abundance are smaller than some of the points. Right panel: Log stellar mass vs. $12+$log(O/H). The stellar mass of Leoncino is based the 3.6~$\micron$ flux, a M/L ratio of 0.47, and the adopted distance of 12.1 Mpc. Typical, star-forming low-mass galaxies with direct-method oxygen abundances and robust distances from the Local Volume Legacy sample form a tight relation in the MZ plane with less dispersion than the LZ relation \citep[][black triangles]{Berg2012}. The best-fitting line is shown as a solid line with the 1$\sigma$ uncertainties shaded in grey. A number of XMP galaxies that were offset in the LZ plane now agree with the MZ relation, including Leoncino, and there is significantly less scatter when considering all the additional XMP galaxies shown. Plot symbols are labeled in the legend with references for the oxygen abundances; a second reference is provided if the stellar mass was obtained from a separate study. See text for details.}
\label{fig:lz_mz}
\vspace{30pt}
\end{figure*}

Finally, we compare Leoncino with other galaxies with oxygen abundances below $\sim5$\% \zsun (12$+$log(O/H)$=$7.35) in the LZ plane, shown in the left panel of Figure~\ref{fig:lz_mz}. As defined in \S\ref{sec:intro}, we refer to these galaxies as ``XMP'' for extremely metal-poor. Again, we limit our comparison sample to those with oxygen abundances measured using the direct method. To simplify the plot, we now show the data from the LVL galaxies in \citet{Berg2012} and the void galaxies from \citet{Pustilnik2016} and \citet{Kniazev2018} as transparent black and cyan points, respectively, with their best-fitting lines and dispersions. 

We include the best-known XMP galaxies, namely I~Zw~18, DDO~68, and SBS 0335-52, and the well-studied XMP galaxies from \citet{Ekta2008}. In addition to being metal poor, these are high surface brightness galaxies and significant outliers in the LVL LZ relation.

We also add more recently discovered XMP galaxies from various studies. Three galaxies are from \citet[][blue right-facing triangles]{Yang2017} who studied a sample of 40 low-mass (log(M$_*$/\msun)$\sim6.5 - 7.5$), compact ($<1$ kpc), high-ionization ([OIII]/[OII] $\sim10 - 60$) galaxies or ``blueberries''.  These galaxies were selected based on properties that are driven by significant recent star formation (i.e., high ionization parameters), which can introduce shifts towards higher luminosity. All three blueberry galaxies are significantly offset from the LZ relation. 

Four galaxies are from \citet{Izotov2018} and \citet[][dark green diamonds]{Izotov2019}, including one system below 12$+$log(O/H) $=$7.0. These systems were found via emission line ratio searches in SDSS and span $\sim5$ mag in luminosity, but only $\sim0.3$ dex in oxygen abundance. Three galaxies are from \citet[][yellow pentagons]{Guseva2017}, who searched SDSS DR12 for metal-poor galaxies that were not included in SDSS DR10.\footnote{We include only direct-method oxygen abundances from \citet{Guseva2017} with a 3$\sigma$ detection of the [O III] $\lambda$4363 line.} These systems all lie off the LZ relation, including two which are among the most extreme outliers in Figure~\ref{fig:lz_mz}.

One galaxy, nicknamed the Little Cub for its location in the constellation Ursa Major \citep[J1044$+$6306;][orange triangle]{Hsyu2017}, lies close to the LZ relation. This system was discovered using different criteria in SDSS. Instead of searching via emission line ratios, \citet{Hsyu2017} used the photometric colors of Leo~P and I~Zw~18 as a template \citep[for similar searches, see also][]{James2015, Hsyu2018}. Finally, two galaxies are from \citet[][light blue circles]{Senchyna2019a}, who identified candidates from SDSS imaging with g$^{\prime}$ and r$^{\prime}$ excesses, corresponding to systems with large specific star formation rates and high equivalent width emission. One of these galaxies, J1005$+$3722, is consistent with the LVL LZ relation.

Note that the absolute luminosities are dependent on the distances to galaxies that are measured using various techniques that have varying accuracies, and sometimes with uncertainties that are not well-quantified. In most studies, the uncertainty on distance is not reported and adds unknown scatter in the LZ relation \citep[see, e.g., ][for an analysis of how high-quality distances reduce the scatter in the LZ relation]{Berg2012}. A few studies report $g$-band luminosities instead of $B$-band, as noted in the legend of Figure~\ref{fig:lz_mz}, which impacts the placement on the LZ relation of order a few tenths of a magnitude. 

With the exception of the low-luminosity galaxies Leo~P and J1005$+$3722, the XMP galaxies are all outliers on the LZ relation. These outlier galaxies share a common set of characteristics, based on the properties reported in the various studies listed in Figure~\ref{fig:lz_mz}. The systems are compact, with physical sizes of their stellar disk of order 1 kpc. Where sufficient photometric depth is achieved, the galaxies also show an extended older stellar population \citep[e.g.,][]{Annibali2013, Sacchi2016}. Many of these galaxies are classified as blue compact dwarfs. Significant populations of massive stars are present in the systems, reported via different metrics including high specific star formation rates, starbursts with young ($\ltsimeq$10 Myr) ages, and high ionization states measured in the gas. Where \hi\ data are available, the galaxies show extended \hi\ disks suggesting high gas-to-stellar mass ratios. The observed gaseous disks ubiquitously show signs of interaction or gravitational disturbance \citep[i.e., I~Zw~18, SBS~0335-052W, DDO~68;][]{vanZee1998a, Ekta2009, Lelli2014c, Cannon2014, Paudel2018, Annibali2019}.

Leoncino has all of these characteristics. It is compact (D $=$ 0.8 kpc) with an extended older population. Leoncino hosts significant recent star formation evidenced by a population of young massive stars (see CMD in Figure~\ref{fig:cmd}). The \hi\ is significantly larger than the stellar distribution (Figure~\ref{fig:hi}) with M$_{\rm HI}$/M$_* = 25$ \msun. While in a void environment, the \hi\ observations are suggestive of a minor interaction with UGC~5186 (see Figure~\ref{fig:hi}). 

The recent star-formation activity in these galaxies contributes to the offset from the LZ relation. If galaxies have high specific star formation rates, their placement in the LZ relation will be biased towards higher luminosity (due to the high luminosity of young massive stars {\it and} the contribution of strong emission lines) when compared to galaxies of similar mass with more typical star-formation activity. Note that the gap in the LZ diagram between the \citet{Berg2012} relation and the XMP delineation is by construction. There are various samples of galaxies that populate this region \citep[e.g.,][]{Berg2016, James2017, Yang2017, Berg2019}. These galaxies are not as extreme in their properties (or as offset from the LZ relation) as XMP galaxies, but these systems may have also experienced an event that lowers the oxygen abundance and triggers star formation.

While comparisons of local star-forming dwarfs found similar dispersions for the LZ and MZ relation \citep{Berg2012}, the LZ relation will be more sensitive to outliers. This is demonstrated in the next section where we compare the oxygen abundances with the stellar mass of galaxies.

\subsection{The Mass-Metallicity Relation}
We present the oxygen abundances as a function of stellar mass in the right panel of Figure~\ref{fig:lz_mz}. Similarly to the LZ relation, we show the subset of the LVL galaxy sample with robustly quantified direct-method abundances and high-quality distances from \citet{Berg2012}. Stellar masses were calculated based on an infrared mass-to-light ratio relation that uses {\it Spitzer Space Telescope} 4.5 $\micron$ and ground-based $K$-band fluxes. The LVL galaxies lie along a tight correlation with a best fit line of 5.61$\pm0.24 + (0.29\pm-0.03) M_*$ and a dispersion in log(O/H) of $\sigma = 0.21$. Going forward, we will refer to this relationship as the LVL MZ relation and it is reproduced on the plot as a solid black line with the $1\sigma$ range shaded in grey. 

Leoncino, shown in red in Figure~\ref{fig:lz_mz}, agrees within the uncertainties with the MZ relation for typical star-forming field galaxies. As described above, we estimated the stellar mass of Leoncino using an infrared flux and assuming a mass-to-light ratio, similar to the approach used in \citet{Berg2012}. Leo~P, shown on the plot as a blue star, also agrees with the MZ relation based on the stellar mass measured from CMD fitting from \citet{McQuinn2015e}. Both systems were discovered via their \hi\ content in the ALFALFA survey and their properties are consistent with an extrapolation from typical low-mass galaxies in the Local Volume. 

Also shown in the MZ diagram in the right panel of Figure~\ref{fig:lz_mz} are star-forming, XMP galaxies from many of the same studies highlighted in the LZ relation in the left panel. Stellar masses were obtained from a variety of sources. Wherever possible, we adopt the stellar masses from the same study from which the oxygen abundances were derived. Specifically, \citet{Yang2017} report stellar masses for the blueberry galaxies based on spectral energy distribution (SED) fits of ugrizy photometry with Starburst99 models \citep{Leitherer1999} after subtracting strong emission lines from the $gri$-band data. \citet{Izotov2018} provide stellar masses using a spectral fitting technique for two galaxies; no stellar masses were available for the two systems in \citet{Izotov2019}. \citet{Guseva2017} report stellar masses but no details were given on their derivation. \citet{Hsyu2017} provide a range in stellar mass for the Little Cub based on the likely range in distance to the galaxy and assuming an optical mass-to-light ratio. \citet{Senchyna2019a} estimate stellar masses with high but unquantified systematic uncertainties using stellar population synthesis modelling and an optical mass-to-light ratio.

For the remaining galaxies, stellar masses were obtained from separate studies. For DDO~68, we use the stellar mass derived from CMD fitting from \citet{Sacchi2016} and assume a 40\% gas return fraction \citep{Vincenzo2016}. For I~Zw~18, we use the lower limit on the stellar mass similarly derived from CMD fitting from \citet{Annibali2013}. For SBS~0335$-$052, we use an estimate of the stellar mass based on modelling the mass and age of stellar clusters in \citet{Schneider2016}, which is noted to have large uncertainties. We obtained the stellar mass for one of the galaxies (UGC~772) in \citet{Ekta2008} from a catalog of interacting dwarfs based on assuming an optical mass-to-light ratio of SDSS $g-$ and $r-$band imaging \citep{Paudel2018}. 

Given the different methods for calculating stellar masses and unquantified systematic uncertainties, the placement of galaxies along the x-axis in the MZ relation shown in Figure~\ref{fig:lz_mz} should only be interpreted as representative. Nonetheless, the change in the distribution of galaxies in the MZ versus the LZ relation is readily apparent. We discuss these differences in detail in \S\ref{sec:pathways}.

\section{Discussion}\label{sec:discuss}
\subsection{Chemical Evolution Pathways for XMP Galaxies}\label{sec:pathways}
Of the XMP galaxies shown in Figure~\ref{fig:lz_mz}, nearly all are outliers on the LVL LZ relation, and some are significantly discrepant. In contrast, when comparing the metallicity with the more fundamental quantity of stellar mass in the MZ plane, a number of XMP galaxies are consistent with the LVL MZ relation and there is significantly less scatter. This suggests two different chemical evolution pathways for XMP galaxies. 

For the outliers in both relations, the lower abundances relative to the expected LZ and MZ trends and the higher luminosities relative to the LZ trend are likely attributable to external events. \citet{Ekta2010b} find evidence of gravitational interactions for metal-poor galaxies, while \citet{Filho2015} suggest there are metal-poor accretion flows in metal-poor galaxies. In both scenarios, the impact of these external events on the galaxies follows the same path. Pristine \hi\ gas is driven into the center regions of the galaxies where it mixes with the ISM, simultaneously diluting the oxygen abundance and fueling a period of high star-formation activity. The common characteristics of high recent star formation activity, high gas-to-star ratios, and disturbed \hi\ discs in cases where interactions may be present, described in detail in \S\ref{sec:lz}, support this interpretation. 

The offset of these same galaxies from the LVL MZ relation supports this scenario, as the extremely low oxygen abundances are incongruent with the expected chemical enrichment at these galaxy masses. A similar conclusion was reached for higher-mass galaxies that are outliers from the mass-metallicity relation \citep{Peeples2009}. 

The larger discrepancies in the LZ plane versus the MZ plane highlight how much of the offset in the LZ relation is due to enhanced luminosity from higher star-formation activity versus the decrease in oxygen abundance by dilution from infall of pristine gas. To explore the potential contribution of these two components, we perform a simple calculation to estimate how much displacement from the LZ relation could be the result of higher luminosities and then we estimate the amount of pristine gas needed to account for the remaining offset. From Figure~\ref{fig:lz_mz}, deviations from the LVL LZ relation for the most extreme cases reach several magnitudes in luminosity and as much as a decade in oxygen abundance. More typical offsets are of order 3$-$4 mag in luminosity and 0.6 dex in oxygen abundance. \citet{Izotov2019} estimate that the luminosity increase for the XMP galaxy J0811$+$4730 could be as much as 2 mags, due to the contribution of the nebular continuum and emission lines to the $g$-band luminosity. A luminosity difference of 2 mags accounts for approximately half the offset from the LZ trend. Thus, we consider how much pristine gas would need to be added to reduce the oxygen abundance by 0.3 dex (e.g., from 12$+$log(O/H) of 7.5 to 7.2), which is consistent with deviations of these galaxies from the LVL MZ relation. Assuming the additional gas does not contain oxygen, the \hi\ mass in the star-forming region where the oxygen abundance is measured would need to be doubled. Based on the existing \hi\ data of XMP galaxies, there are substantial gas reservoirs in these systems that could provide this additional material \citep[e.g., ][]{Filho2013}. 

In summary, for the outliers in both the LZ and MZ planes, the pathway to becoming such a metal-poor galaxy requires {\em an external event}, and {\em the offset from the LZ and MZ relations is temporary}. The ISM is expected to be quickly enriched with newly synthesized material from the high star-formation activity, and, after the high star formation declines, the high luminosity will subsequently decline in turn. The enrichment of the ISM will also move the galaxies towards the LZ and MZ relations. 

In contrast, for the galaxies that are outliers in the LZ relation but {\it agree} with the MZ relation, including Leoncino, secular evolution may be the dominant factor in determining the low metal content of XMP galaxies. The displacement in the LZ plane may be due to higher levels of recent star formation, but the low oxygen abundance is likely a result primarily of inefficient star formation and metal loss via stellar-feedback driven galactic winds. This scenario was explored in detail for Leo~P, one of only two XMP galaxies that agree with both the LVL LZ and MZ relations, whose low oxygen abundance has been attributed to low star formation rates and metal expulsion \citep{McQuinn2015f}. 

Specifically for Leoncino, the offset from the LZ trend is attributable to the population of young, massive stars whose formation may have been triggered by a minor interaction with UGC~5186. Using the resolved stars in the CMD in Figure~\ref{fig:cmd}, we estimate Leoncino would be $\sim1.3$ mag fainter in the $V$-band equivalent F606W filter without the population of bright upper main sequence stars. This luminosity offset if derived by accounting for the contribution from stars brighter than 27.0 mag in the F606W filter and bluer than 0.35 mag in F606W$-$F814W colors, corresponding to the 100 Myr isochrone from the PARSEC stellar evolution models. A simple modellng exercise with {\sc Starburst99} \citep{Leitherer1999} shows this corresponds to a 100 Myr burst of star formation of $\sim$5\% the stellar mass of Leoncino. Assuming a $B-V$ color of $\sim0$ for these hot stars, we apply this 1.3 mag difference to the integrated $B$-band luminosity and estimate $M_B \sim -9.3$ mag for the remainder of the galaxy. At this fainter magnitude, Leoncino would agree within the uncertainties with the LZ relation of typical star-forming galaxies in the Local Volume.

Thus, recent star-formation activity, possibly triggered by an interaction with UGC~5186, is likely responsible for driving Leoncino off the LVL LZ relation. Modest gas infall from the outer disk of Leoncino may also contribute to the offset from the LZ relation to lower abundances (and it may also have played a role in the case of the other galaxies that agree with the MZ relation); however, if infall is occurring, it is likely limited in scope. In summary, an external event does not appear to be the primary driver of the low gas-phase oxygen abundance in Leoncino. Instead, the main chemical evolution pathway is secular in nature and consistent with an extrapolation of typical star-forming galaxies to lower masses.

\subsection{The Missing XMP Galaxies on the LZ Relation}
Searches for metal-poor galaxies are becoming more fruitful and we are beginning to populate the low-metallicity end of the LZ relation with robust, direct-method oxygen abundances (see \S\ref{sec:intro}). Despite these successes, detections of XMP galaxies with abundances less than $\sim$5\% \zsun\ remain rare. When systems are identified, they are almost always outliers in the LZ relation; Leo~P and J1005$+$3722 are the only two star-forming galaxies with a direct oxygen abundance below 5\% \zsun\ that are consistent with the relation. Even when searches are designed for low-luminosity XMP galaxies \citep[i.e., using search criteria modeled after the color and morphology of the low-luminosity XMP galaxy Leo~P;][]{James2015, Hsyu2017, Hsyu2018}, the XMP systems they identify are offset from the LZ relation. 

The disparity between finding galaxies with oxygen abundances below 5\% \zsun\ that lie off or on the LZ trend is likely caused by three factors. The first is a selection effect. XMP galaxies that lie along the LZ relation have appreciably fainter luminosities and lower surface brightnesses and are more difficult to detect than those that lie off the relation. Starbursting XMP galaxies can be detected out to vastly greater distances (e.g., at z $\sim$ 0.2, as opposed to the Local Volume), so that comparing their intrinsic volume number densities is very difficult. Exacerbating the disparity, the properties of the higher luminosity XMP systems $-$ such as high ionization states and strong emission lines that dominate their colors $-$ facilitate unique search criteria which further increase the likelihood of detection in SDSS data relative to quiescent XMP galaxies\footnote{We emphasize, however, that XMP galaxies which lie off the LZ relation are still a challenging population to find.} \citep[e.g.,][]{Yang2017, Berg2016}. 

The second factor is that there may be very metal-poor galaxies in current surveys that cannot be identified as such because their gas-phase oxygen abundance measurements are unattainable. In order to measure an oxygen abundance in a galaxy, there must be an \hii\ region driven by a high-mass O star. At low star formation rates corresponding to the low-luminosity end of the LZ relation ($\sim10^{-5}$ \msun\ yr$^{-1}$), the upper IMF is not always fully populated and, thus, there may be a population of XMP galaxies without bright \hii\ regions, or, in fact, any \hii\ regions at all. It is likely that there are additional low-luminosity XMP galaxies in current surveys that are unclassified as such. 

A prototype for this class of uncatalogued low-luminosity XMP galaxies could be the Local Group Sagittarius Dwarf Irregular galaxy. \citet{Skillman1989b} measured a nebular oxygen abundance of 12 $+$ log(O/H) $=$ 7.4, just above our 5\% \zsun\ threshold, but this was estimated using the R$_{23}$ strong line method because the low surface brightness nature of the \hii\ region did not allow an [\ion{O}{3}] $\lambda$4363 measurement. A similar result was obtained by \citet{Saviane2002}. With this oxygen abundance, and the luminosity and stellar mass reported in \citet{Skillman1989b} and \citet{Saviane2002} respectively, Sag DIG is consistent with the LZ and MZ relationships. Although this galaxy is left out of most XMP compilations because of the lack of an [\ion{O}{3}] $\lambda$4363 detection, it is sufficiently nearby that $HST$ observations of the stellar populations are able to show that even the recently formed stars are substantially metal-poor \citep[Z $ \approx$ 0.0004;][]{Momany2005}. Similarly, the Local Group galaxy LGS-3 has been forming stars right up to the present, and the stellar metallicity is consistent with the XMP criteria of $\sim$5\% \zsun, but the galaxy has no \hii\ regions at all \citep{Hidalgo2011}.

Finally, the number of observable metal-poor galaxies may be low given the apparent magnitude limits of surveys and a possible turn-over in the galaxy luminosity function. Detailed modeling of the galaxy luminosity function at low masses predicts of order six quiescent galaxies with oxygen abundances below $\sim$10\% \zsun\ from the SDSS-DR7 spectroscopic survey when assuming that the baryon fraction decreases in low-mass dark matter halos \citep{SanchezAlmeida2017}. In this scenario, the galaxy luminosity function is not extrapolated to rise at lower masses, but flattens as low-mass halos are quenched due to heating from the ultraviolet background radiation and stellar feedback. With suppressed numbers of low-mass systems, the prediction is that there will be a correspondingly lower number of XMP systems in the volume probed by current surveys. The dominant restricting factor is shown to be the apparent magnitude limit, suggesting that upcoming, deep, wide field surveys have the potential for discovering significant numbers of XMP galaxies. In the meantime, low-luminosity XMP galaxies in the nearby universe that lie on the LVL LZ relation remain elusive. 

\section{Conclusions}\label{sec:conclusions}
The number of discoveries of extremely metal-poor (XMP; 12$+$log(O/H)$<7.35$ or $\ltsimeq \frac{1}{20}$ \zsun) galaxies in the past few years is impressive and offers a growing sample that can be used to study star formation in ISM conditions analogous to the early universe. The Leoncino dwarf (AGC~198691), discovered via its \hi\ in the ALFALFA survey, is one of these newly found systems, with a gas-phase oxygen abundance of $12+$log(O/H)$=7.12\pm0.04$ measured via the temperature sensitive  [\ion{O}{3}] $\lambda$4363 auroral line (i.e., the direct method). Using $HST$ optical imaging of the resolved stellar populations of Leoncino, we find that the CMD is sparsely populated, with only 147 sources brighter than F814W$\sim28.5$ mag. Despite the low star counts, the galaxy has a well-defined main sequence branch of young, massive stars. By identifying the TRGB, the distance is measured to be 12.1$^{+1.7}_{-3.4}$ Mpc, placing Leoncino in an under-dense galactic environment and in the nearby Void No.~12 \citep{Pustilnik2019}. Even though the overall density of galaxies is low, Leoncino is located at a projected distance of only 46 kpc from UGC~5186 and an offset of only 35 km s$^{-1}$ in radial velocity and shows signs of an interaction including a putative detection of \hi\ between the two galaxies (see Figure~\ref{fig:hi}; J. M. Cannon et al., in preparation).

Similar to nearly all of the known XMP galaxies in the nearby universe, Leoncino lies off the luminosity-metallicity (LZ) relation defined by gas-rich, star-forming galaxies in the Local Volume \citep{Berg2012}. We limit our comparison to only systems with oxygen abundances measured with the direct method to avoid introducing systematic uncertainties from different metallicity indicators (see Figure~\ref{fig:lz_berg_voids}). When we constrain that sample in this way, we find no significant differences in low-metallicity galaxies found in void environments from those evolving in more typical field environments. Additional direct-method oxygen abundances would be useful in confirming this result. Thus, even though Leoncino is located in a void, the offset from the LZ relation is not likely due to a distinct evolutionary path in a low-density environment.

The XMP galaxies that are outliers from the LZ relation, including Leoncino, share a number of common characteristics including a compact size, significant recent star-formation activity, and, where measured, extended and disturbed \hi\ disks indicative of interactions. When examined in the MZ plane, which involves the more fundamental quantity of stellar mass, only a subset of the galaxies remain outliers and there is significantly less scatter (Figure~\ref{fig:lz_mz}). For the galaxies that remain outliers, we find their properties to be consistent with previous interpretations that the low oxygen abundances and high luminosities of XMP galaxies offset from the LZ relation are due to the infall of pristine gas, predominantly from the outer disks of of the galaxies themselves driven by an interaction \citep{Ekta2010b, Peeples2009} or by metal-poor gas accretion \citep{Filho2015}. Although intrinsically rare, these galaxies are preferentially found in emission-line surveys and high surface brightness surveys (e.g., SDSS) and can be found to very large distances (i.e., over larger volumes).

The few galaxies that agree with the MZ relation $-$ including Leoncino $-$ have low oxygen abundances that are consistent with expectations from secular evolution. Their chemical evolution can be explained based primarily on inefficient star formation and metal-loss via stellar-feedback driven galactic winds at these low galaxy masses. The offset from the LZ relation for these systems is likely still due to a recent burst of star formation. In the case of Leoncino, the recent star formation appears triggered by a minor interaction with UGC~5186. If the young, upper main sequence stars in the CMD are excluded, Leoncino agrees with the LZ relation based on the fainter, integrated luminosity. Similar to the low-luminosity galaxy Leo~P that is consistent with both the LVL LZ and MZ relations, Leoncino was discovered via its \hi\ content from the ALFALFA survey, which appears to be an efficient approach to finding XMP galaxies whose low oxygen abundance is a result of secular evolution at the low-mass end of the galaxy mass function. Upcoming, blind \hi\ surveys open up promising opportunities for increasing detections of such systems.

XMP galaxies that are not outliers to the LZ relation remain difficult to find. This is likely due to a number of factors including (i) a bias against finding ever-fainter galaxies compared to the higher surface brightness XMP galaxies, (ii) the inability of measuring oxygen abundances in all low-luminosity galaxies as they do not ubiquitously host massive O stars and the \hii\ regions needed for abundance work, and (iii) the possible suppression of the number of low-mass galaxies (i.e., lower galaxy counts at the turn-over of the luminosity function) combined with apparent magnitude limits of current surveys \citep{SanchezAlmeida2017}. 

\section{Acknowledgments}
Support for program HST-GO-15243 (PI McQuinn) was provided by NASA through a grant from the Space Telescope Science Institute, which is operated by the Associations of Universities for Research in Astronomy, Incorporated, under NASA contract NAS5- 26555. KBWM is also supported by Rutgers University, NSF grant AST-1806926, and NASA grant HST-GO-15227. MPH is supported by grants NSF/AST-1714828 and from the Brinson Foundation. JMC and MK are supported by Macalester College. KLR is supported by NSF grant AST-1615483. This research has made use of NASA’s Astrophysics Data System Bibliographic Services and the NASA/IPAC Extragalactic Database (NED), which is operated by the Jet Propulsion Laboratory, California Institute of Technology, under contract with the National Aeronautics and Space Administration.

\facility{Hubble Space Telescope}

\renewcommand\bibname{{References}}
\bibliography{../../bibliography.bib}
\end{document}